\documentclass[a4paper,fleqn,usenatbib]{mnras}

\usepackage[T1]{fontenc}
\usepackage{ae,aecompl}

\usepackage{graphicx}	
\usepackage{epstopdf}
\usepackage{savesym}
\usepackage{amsmath}	
\usepackage{amssymb}	
\savesymbol{iint}
\usepackage{txfonts}
\restoresymbol{TXF}{iint}

\usepackage{caption}
\usepackage{subcaption}

\def\apjs {ApJS}
\def\apjl {ApJL}

\def\aap {A\&A}

\DeclareRobustCommand{\ion}[2]{%
\relax\ifmmode
\ifx\testbx\f@series
{\mathbf{#1\,\mathsc{#2}}}\else
{\mathrm{#1\,\mathsc{#2}}}\fi
\else\textup{#1\,{\mdseries\textsc{#2}}}%
\fi}

\title[GAMA: AGN in groups]{Galaxy And Mass Assembly (GAMA): The effect of galaxy group environment on active galactic nuclei}

\author[Y. A. Gordon et al.]{Yjan A. Gordon,$^{1}$\thanks{E-mail: y.gordon@2014.hull.ac.uk (YAG)}
Kevin A. Pimbblet,$^{1}$
Matt S. Owers,$^{2, 3}$
Joss Bland-Hawthorn$^{4}$
\newauthor Sarah Brough,$^{5}$
Michael J. I. Brown,$^{6}$
Michelle E. Cluver,$^{7}$
Scott M. Croom,$^{4}$
\newauthor Benne W. Holwerda,$^{8, 9}$
Jonathan Loveday,$^{10}$
Smriti Mahajan$^{11}$
and Lingyu Wang$^{12, 13}$
\\
$^{1}$E. A. Milne Centre for Astrophysics, University of Hull, Cottingham Road, Kingston-upon-Hull HU6 7RX, U.K.\\
$^{2}$Department of Physics and Astronomy, Macquarie University, NSW 2109, Autralia\\
$^{3}$Australian Astronomical Observatory, 105 Dehli Road, North Ryde, NSW 2113, Australia\\
$^{4}$Sydney Institute for Astronomy (SIfA), School of Physics, The University of Sydney, NSW2006, Australia\\
$^{5}$School of Physics, University of New South Wales, NSW2052, Australia\\
$^{6}$School of Physics and Astronomy, Monash University, Clayton, Victoria 3800, Australia\\
$^{7}$Department of Physics and Astronomy, University of the Western Cape, Robert Sobukwe Road, Bellville, 7535, South Africa\\
$^{8}$Department of Physics and Astronomy, University of Louisville, Louisville KY 40292, U.S.A.\\
$^{9}$University of Leiden, Sterrenwacht Leiden, Niels Bohrweg 2, NL-2333 CA Leiden, The Netherlands\\
$^{10}$Astronomy Centre, University of Sussex, Falmer, Brighton BN1 9QH, U.K.\\
$^{11}$Indian Institute of Science Education and Research Mohali (IISERM), Knowledge City, Sector 81, SAS Nagar, Manauli PO 140306, India\\
$^{12}$SRON Netherlands Institute for Space Research, Landleven 12, 9747 AD, Groningen, The Netherlands\\
$^{13}$Kapteyn Astronomical Institute, University of Groningen, Postbus 800, 9700 AV, Groningen, The Netherlands
}

\date{Accepted 2018 January 10. Received 2018 January 3; in original form 2017 December 8}

\pubyear{2017}

\begin{document}
\label{firstpage}
\pagerange{\pageref{firstpage}--\pageref{lastpage}}
\maketitle

\begin{abstract}
In galaxy clusters, efficiently accreting active galactic nuclei (AGN) are preferentially located in the infall regions of the cluster projected phase-space, and are rarely found in the cluster core.
This has been attributed to both an increase in triggering opportunities for infalling galaxies, and a reduction of those mechanisms in the hot, virialised, cluster core.
Exploiting the depth and completeness ($98\,$per cent at $r<19.8\,$mag) of the Galaxy And Mass Assembly survey (GAMA), we probe down the group halo mass function to assess whether AGN are found in the same regions in groups as they are in clusters.
We select 451 optical AGN from 7498 galaxies with $\log_{10}(M_*/\text{M}_\odot) > 9.9$ in 695 groups with $11.53\leq \log_{10}(M_{200}/\text{M}_\odot) \leq 14.56$ at $z<0.15$.
By analysing the projected phase-space positions of these galaxies we demonstrate that when split both radially, and into physically derived infalling and core populations, AGN position within group projected phase-space is dependent on halo mass.
For groups with $\log_{10}(M_{200}/\text{M}_\odot)>13.5$, AGN are preferentially found in the infalling galaxy population with $3.6\sigma$ confidence.
At lower halo masses we observe no difference in AGN fraction between core and infalling galaxies.
These observations support a model where a reduced number of low-speed interactions, ram pressure stripping and intra-group/cluster medium temperature, the dominance of which increase with halo mass, work to  inhibit AGN in the cores of groups and clusters with $\log_{10}(M_{200}/\text{M}_\odot)>13.5$, but do not significantly affect nuclear activity in cores of less massive structures.
\end{abstract}




\section{Introduction}
\label{agngroupsintro}
Active galactic nuclei (AGN) are powered by the active accretion of matter onto the central super-massive black hole (SMBH) of a galaxy. 
Consequently, AGN may be triggered by mechanisms that introduce a new supply of cold gas, i.e., mergers \citep{Sanders1988, Krongold2002}, that can then act as a fuel supply for nuclear activity.
Alternatively, physical mechanisms that have the potential to destabilise the cold gas reservoirs already within a galaxy, such as harassment \citep{Moore1996} or ram pressure stripping \citep[RPS,][]{Poggianti2017, Marshall2017}, may trigger an infall of this gas toward the nucleus, where it can then be accreted by the SMBH.

Determining if these mechanisms actually do trigger AGN can be achieved by environmental analysis of the AGN host galaxies. The incidence of AGN has been shown to be enhanced in galaxies in very close pairs, and thus likely in the process of merging, relative to those galaxies not in a pair \citep{Alonso2007, Woods2007, Ellison2011}. Galaxy harassment and RPS occur within the denser environment of galaxy clusters. Here the infall of a galaxy toward the bottom of the gravitational potential well can both increase the number of high velocity close encounters a galaxy has \citep{Moore1996}, and subject the galaxy to the pressure of intra-cluster medium (ICM).

In the cluster environment at low-intermediate redshift, studies have shown that efficiently accreting AGN, i.e., not the radio-mode dominated AGN with intrinsically lower accretion rates \citep[][]{Hardcastle2007, Best2012}, preferentially inhabit regions of cluster projected phase-space associated with the infalling population \citep[e.g.,][]{Ruderman2005, Haines2012, Pimbblet2013, Pentericci2013, Ehlert2013}. That is to say AGN are found amongst the cluster population experiencing harassment and RPS for the first time \citep[][]{Mahajan2012}.

In contrast to the infall region, the cluster cores have generally been observed to be relatively barren of AGN (\citealp{Gilmour2007, Gavazzi2011, Pimbblet2012}, cf., \citealp{Ruderman2005}). That is not to say that AGN aren't found at all in cluster cores, indeed a small fraction of brightest cluster galaxies are known to harbour AGN \citep[e.g.,][]{Best2007, Fraser-McKelvie2014, Green2016}, but that they are relatively rare in comparison to the infall regions of the structure.
This dearth of AGN in cluster cores may be the result of nuclear activity having run its course during the time it takes for the galaxy to fall into the cluster centre. Alternatively, it may be the case that the environment of the cluster core is unfavourable to AGN. The cores of relaxed massive clusters have had time to virialise, and consequently the galaxy interactions in this region are high speed in nature and less conducive to galaxy mergers. The resultant lack of low-speed galaxy-galaxy interactions \citep{Ostriker1980} may prevent AGN from being triggered in this region. Furthermore, the ram pressure experienced by galaxies close to the cluster centre may be too high to trigger an AGN \citep{Marshall2017}, instead stripping the galaxy of its gas. This may present observationally as one-sided tails \citep[e.g.][]{Kenney2004, Fumagalli2014} or `jellyfish' galaxies should star formation occur in those tails \citep[e.g.][]{Owers2012, Ebeling2014}. Such stripping would eventually starve an AGN of a potential fuel supply. A further mechanism which may inhibit nuclear activity in cluster cores is the temperature of the ICM. As the ICM may be of the order of tens of megaKelvin \citep{Fabian1994}, accretion onto galaxies is unlikely \citep{Davies2017}. In combination, these mechanisms result in galaxies in the cluster core that have an intrinsically low reservoir of cold gas \citep{Giovanelli1985} and thus cannot easily fuel an AGN.

Whilst these effects are well established in clusters, they may not extend to groups. Unlike clusters, which may contain many thousands of galaxies, groups have only up to a few tens of members, and significantly lower halo masses. The dynamics of the group environment may permit more galaxy-galaxy interactions that are effective in driving gas toward the galactic nucleus within in the group centre. Furthermore, the lower halo masses of groups will result in a smaller heating effect from the virial shock acting on the intra-group medium \citep[IGM,][]{Grootes2017}. This may allow for easier, or more rapid, accretion of the IGM onto galaxies and act as a potential fuel reservoir. Finally, the lower density of the galaxy groups results in lower ram pressures affecting infalling galaxies \citep{Marshall2017}. Consequently RPS will be less likely to strip a galaxy of a large fraction of its gas. Ergo, these environmental differences between group and cluster galaxies may foster the presence of AGN in the cores of groups. 

While galaxy groups have been shown to have a higher global AGN fraction than clusters \citep{Shen2007, Arnold2009, Tzanavaris2014, Oh2014}, it is unestablished whether the effects of position in projected phase-space seen in clusters are present or absent in groups.
Where efforts have been made to study the effect of groups on AGN these studies are limited by small numbers of detected groups. This hinders the ability to stack groups to perform a phase-space analysis with any level of statistical confidence.
The Galaxy And Mass Assembly survey \citep[GAMA,][]{Driver2011, Liske2015} is highly spectroscopically complete ($98\,$per cent at $r<19.8\,$mag, \citealp{Liske2015}) making it well suited to environmental analyses of galaxies \citep[e.g.,][]{Brough2013, Casteels2014, Robotham2014, Alpaslan2015, Davies2016, Ching2017, Gordon2017, Barsanti2017}. GAMA is thus ideal for studying the lower halo mass regime of galaxy groups and has detected more than 23000 groups with 2 -- 316 members \citep[][]{Robotham2011}.

In this work we investigate the effect of the group environment on galactic nuclear activity by using spectroscopically selected AGN and version 9 of the GAMA group catalogue.
We analyse prevalence of AGN as a function of group mass and position in group projected phase-space. For our analysis we only use groups with at least 5 members (full details are given in Section \ref{SSG3C}). We are thus expanding on the work of \citet{Pimbblet2013} and probing further down the group halo mass function, and, as a result of the depth of GAMA, to lower galactic stellar masses.
 
In Section \ref{Sdata} we detail the GAMA survey and the specific data used in this work. Our results are given in Section \ref{Sobs}, and our discussion is presented in Section \ref{Sdiscuss}. Section \ref{Sconc} is a summary of our conclusions. Throughout this paper a standard flat $\Lambda$CDM cosmology is assumed with $h=0.75$, $H_{0}=100h\,\text{km}\,\text{s}^{-1}\,\text{Mpc}^{-1}$, $\Omega_{M}=0.25$ and $\Omega_{\Lambda}=0.75$.

\section{Data}
\label{Sdata}

\subsection{Galaxy And Mass Assembly}
\label{SSGAMA}
The GAMA survey spectroscopic campaign was undertaken with the Anglo-Australian telescope (AAT) at Siding Spring Observatory between 2008 and 2014 \citep{Driver2011, Liske2015}.
During this period, spectra with a resolution of $R\approx 1300$ covering the wavelength range $3750-8850\,\text{\AA}$ were obtained using the 2dF/AAOmega spectrograph for over 250,000 galaxies \citep{Hopkins2013}.
The GAMA survey footprint covers 5 regions of the sky totalling $286\,\text{deg}^{2}$ with extremely high spectroscopic completeness. Specifically, at $r<19.8\,$mag, completeness is $> 98\,$per cent in the three equatorial regions (G09, G12, G15) which we use in this work. 

\subsection{Galaxy groups}
\label{SSG3C}
\subsubsection{GAMA Group Catalogue, {\textsc{G$^3$Cv09}}}
The depth and completeness of GAMA naturally lends itself to the reliable detection of galaxy groups of lower mass than would be possible with a shallower and less spectroscopically complete survey. The resultant GAMA group catalogue is constructed through the use of a friends-of-friends (FoF) algorithm \citep{Robotham2011}. We use the latest version of the group catalogue, G$^3$Cv09, which covers the three GAMA equatorial regions, and contains $75029$ galaxies in $23654$ groups with $z<0.6$. The number of group members, $N_{\text{FoF}}$, ranges from 2 ($14876$ groups) to 316 \citep[Abell 1882,][]{Owers2013}.

We select the groups for use in this work to be within a suitable redshift so that we can be confident in our mass completeness, and having enough group members such that low group multiplicity does not bias our projected phase-space analysis. To determine a redshift limit to use we use the distribution of maximum observable redshift of the GAMA galaxies at $r<19.8\,$mag, as calculated by \citet{Taylor2011}.
By limiting our analysis to $z<0.15$, we are able to probe down to a stellar mass of $10^{9.9}\,\text{M}_{\odot}$ with $90\,$per cent completeness.

\subsubsection{Group membership}
\label{S2groupmem}
The FoF algorithm used to determine the group membership in {\textsc{G$^3$Cv09}} is more reliable at detecting low mass groups than halo-galaxy grouping methods \citep{Robotham2011}. However, this method preferentially detects group members at low projected radii \citep{Barsanti2017}, and hence will bias the results of any projected phase-space analysis conducted. To counter this, we define our own group sample membership, using the GAMA group catalogue as a starting point. Using the observed group velocity dispersion, $\sigma_{\text{group}}$, we calculate $R_{200}$ as per \citep{Carlberg1997},
\begin{equation}
R_{200} = \frac{\sqrt{3}\sigma_{\text{group}}}{10H(z)},
\end{equation}
and the halo masses, $M_{200}$, using
\begin{equation}
M_{200} = \frac{\sigma_{\text{group}}^{3}}{1090^{3}h(z)}10^{15}\,\text{M}_\odot
\label{m200eqn}
\end{equation}
derived by \citet{Munari2013}.
Using the central galaxy from the respective FoF group as our group centre, we then count galaxies within $3.5\sigma_{\text{group}}$ and $3.5R_{200}$, comparable to the analysis conducted by \citet{Pimbblet2013}, as group members.
When compared to mock catalogues, derived group parameters are shown to be more reliable for GAMA groups with $N_{\text{FoF}} \geq 5$ \citep{Robotham2011} suggesting that these groups are more representative of the expected group population. We thus use only these groups as the seeds for our group sample.

At high projected separation and velocities, interloping field galaxies that are serendipitously selected in the group may be a source of substantial sample contamination \citep{Rines2005, Oman2016, Owers2017}. To counter this, we further require that galaxies satisfy the infall criteria of spherical Navarro-Frenk-White \citep[NFW,][]{Navarro1996} profiles as per the method used by \citet{Barsanti2017}. That is, 
\begin{multline}
V_{\text{infall}}(R/R_{200}) <\\ V_{200}\sqrt{\frac{2}{R/R_{200}}\frac{\ln(1+\kappa R/R_{200})-(\kappa R/R_{200})/(1+\kappa R/R_{200})}{\ln(1+\kappa)-\kappa/(1+\kappa)}
}
\end{multline}
where $\kappa$ is the \citet{Dolag2004} halo concentration index given by
\begin{equation}
\kappa(M_{200}, z) = \frac{9.59}{1+z_\text{group}}\bigg(\frac{M_{200}}{10^{14}\,\text{M}_\odot\,h^{-1}}\bigg)^{-0.102}
\end{equation}
$z_\text{group}$ is the median redshift of the group, and \text{$V_{200} = \sqrt{\text{G}M_{200}/R_{200}}$}.

One potential complication of this method is the potential for galaxies being assigned to multiple groups. In this event, a galaxy is assigned to the group that provides the smallest value for the parameter $C$ \citep{Smith2004}. $C$ is given by
\begin{equation}
C = \frac{(cz_{\text{gal}} - cz_{\text{group}})^{2}}{\sigma_{\text{group}}^{2}} - 4\log_{10}\Bigg(1-\frac{R}{R_{\text{group}}}\Bigg)
\end{equation}
where $R$ is the projected separation of the galaxy from the group centre, and $R_\text{group}$ is $3.5R_{200}$.
We further require that at least 5 members are assigned to each group via this method to be counted in our projected phase-space analysis, resulting in a sample of 723 groups at $z_{\text{group}} <0.15$.

\subsubsection{Group substructure}
For high-multiplicity groups, we do not wish to bias our results due to any significant substructure that may be present in the group core. To ensure this is not the case, we subject large groups to a $\Delta$, or DS \citep[][]{Dressler1988} test for subclustering. Note that it is the centres of groups in which we are concerned about substructure. To find substructure here would indicate that the group is not a virialised structure and hence unsuitable for our analysis of projected phase-space as this may affect the likelihood of AGN to be found in this region. Moreover, such central substructure may influence the measured dynamical mass of a group.
The DS test is sensitive to substructure in group sizes as low as 30 members. We therefore perform this test on  groups with at least 30 members within $R_{200}$ of the group centre. Of our 723 groups, 41 satisfy this criterion.

To apply the DS test, for each galaxy, $i$, the $n$ nearest neighbours are used to determine a local mean velocity and velocity dispersion. These are then compared to the group global values and a statistic, $\delta$, can be calculated by:
\begin{equation}
\delta_{i}^{2} = \frac{n+1}{\sigma_{\text{global}}} [(v_{\text{local}} - v_{\text{global}})^{2} + (\sigma_{\text{local}} - \sigma_{\text{global}})^{2}]
\end{equation}
where $v$ is the mean velocity and $\sigma$ the velocity dispersion. For each group a cumulative statistic, $\Delta$ is obtained by $\Delta = \Sigma\delta_i$. \citet{Pinkney1996} and \citet{Einasto2012} recommend that the number of nearest neighbours, $n$, be the square root of the number of galaxies used to determine the the cumulative statistic, $\Delta$, i.e., $n = \sqrt{N_{R<R_{200}}}$.
A $p$-value for the absence of substructure is obtained by calibrating $\Delta$ by the use of Monte Carlo simulations whereby $p = N(\Delta_{\text{MC}}>\Delta_{\text{obs}})/N(\text{MC})$. Such simulations are performed by randomly shuffling the galaxy velocities, therefore removing any locally coherent velocity substructure information that is the result of dynamical substructure within the group. For our DS tests we run 1000 Monte Carlo simulations. Given this, and the sensitivity of the DS test to substructure of the order $1/7$ of the total cluster mass \citep{Pinkney1996}, a low $p$-value is indicative of significant substructure.
Of the 41 groups we subjected to the DS test, 18 resulted in $p<0.01$, and were hence eliminated from the analysis.

\begin{figure}
	\centering
	\begin{subfigure}{\columnwidth}
	\includegraphics[width=\columnwidth]{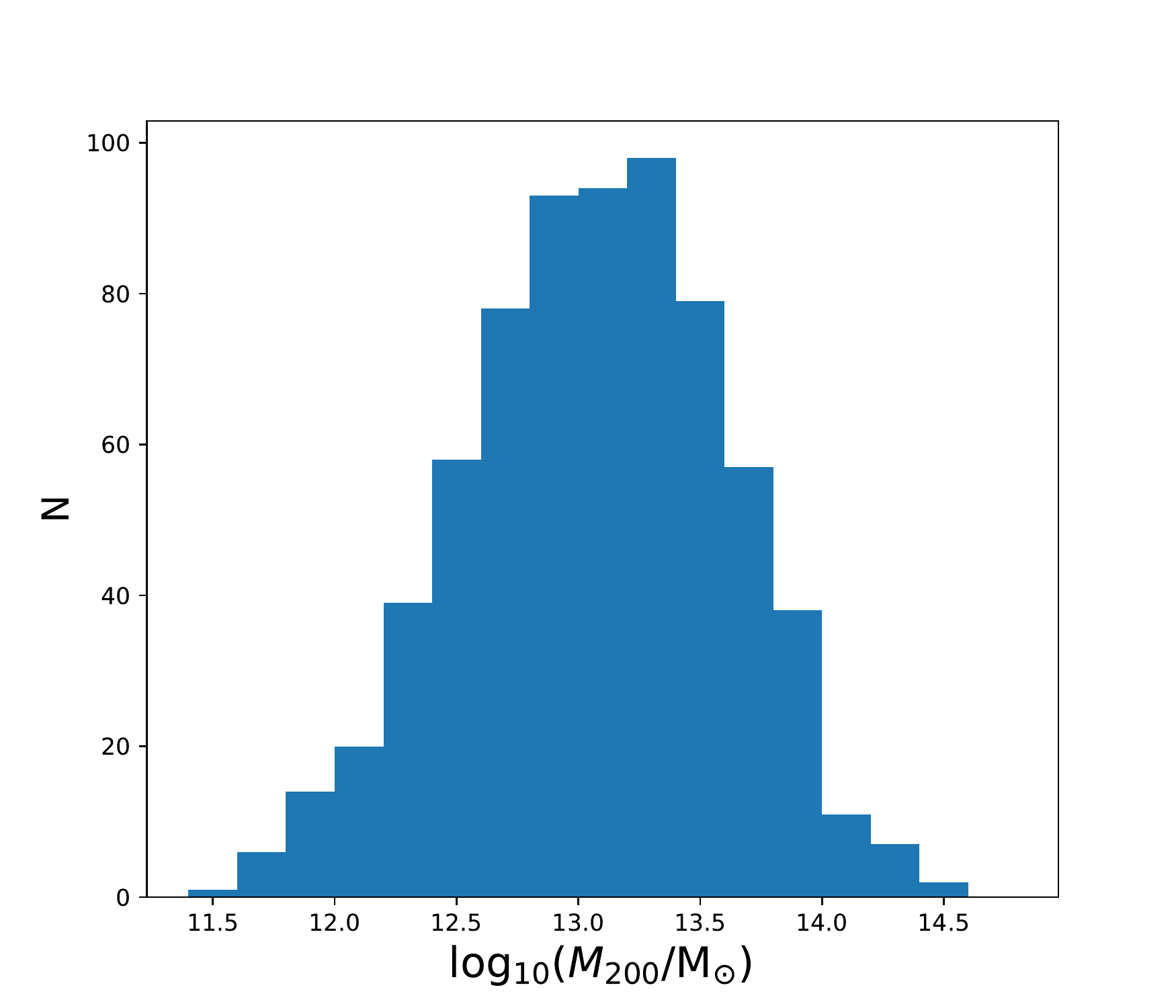}
	\caption[]{}
	\end{subfigure}
	\begin{subfigure}{\columnwidth}
	\includegraphics[width=\columnwidth]{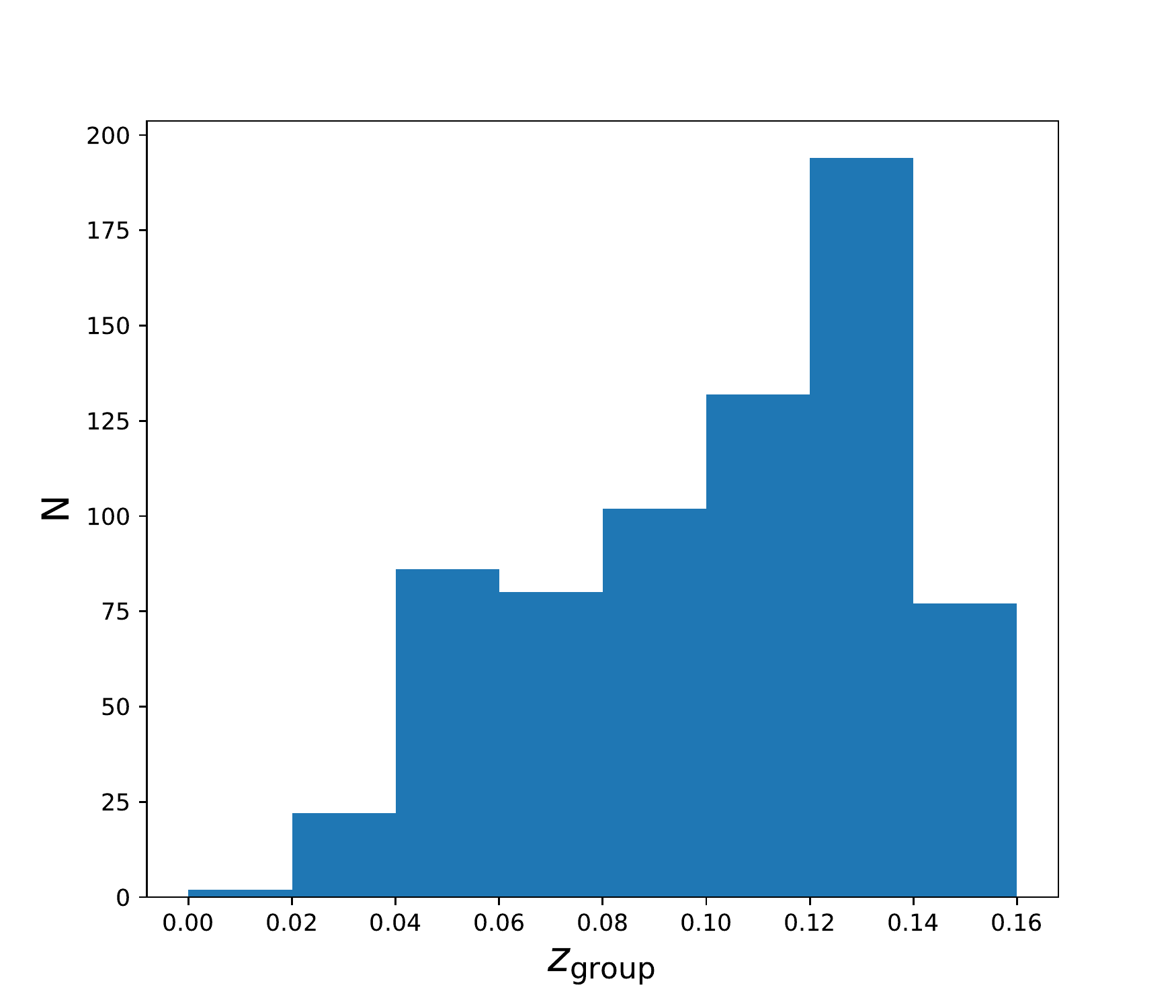}
	\caption[]{}
	\end{subfigure}
\caption{The distributions of the halo mass ($M_{200}$, a) and median redshift (b) of our group sample.}
\label{groupdists}
\end{figure}

This leaves us with 7498 galaxies with $M_* > 10^{9.9}\,\text{M}_\odot$ in 695 groups of between 5 and 418 members, the distributions of the group multiplicity, velocity dispersions, and halo masses of this sample are described in Table \ref{Tgroup}, and the halo mass and redshift distributions are shown in full in Figure \ref{groupdists}.

\begin{table*}
\centering
\caption{The number of group members($N$), group velocity dispersion ($\sigma_{\text{group}}$), radius ($R_{200}$), and mass($M_{200}$) of the groups used for the projected phase-space analysis in this paper. For each property, the mean, standard deviation and the 0th, 25th, 50th, 75th, and 100th percentiles are given.}
\begin{tabular}{lccccccc}
\hline
Group property & $\mu$ & $\sigma$ & Min & $Q_1$ & $Q_2$ & $Q_3$ & Max\\
\hline
$N$ & $21.49$ & $26.89$ & $5.00$ & $9.00$ & $15.00$ & $25.00$ & $418.00$\\
$\sigma_{\text{group}}$[km/s] & $248$ & $103$ & $69$ & $173$ & $228$ & $304$ & $718$\\
$R_{200}$ [kpc/$h$] & $548$ & $226$ & $159$ & $381$ & $504$ & $675$ & $1567$\\
$\log_{10} (M_{200}/\text{M}_{\odot})$ & $13.06$ & $0.54$ & $11.53$ & $12.71$ & $13.07$ & $13.45$ & $14.56$\\
\hline
\end{tabular}
\label{Tgroup}
\end{table*}

\subsection{AGN selection}
\label{SSAGN}
\subsubsection{A complete AGN sample}
The GAMA AGN  sample of \citet{Gordon2017} contains 954 broad- and narrow-line AGN, relying on the \citet{Kewley2001} criteria,
\begin{equation}
\log_{10}\Bigg(\frac{\text{[\ion{O}{iii}]~$\lambda5007$}}{\text{H}\beta}\Bigg) > \frac{0.61}{\log_{10}\Big(\frac{\text{[\ion{N}{ii}]~$\lambda6583$}}{\text{H}\alpha}\Big)-0.47} + 1.19
\label{EQKe01}
\end{equation}
to classify narrow-line AGN on a Baldwin-Phillips-Terlevich \citep[BPT,][]{Baldwin1981} diagnostic diagram. This requires the detection of four emission lines, H$\beta$, [\ion{O}{iii}]~$\lambda 5007$, H$\alpha$, [\ion{N}{ii}]~$\lambda 6583$ at $\text{S/N} >3$, and, consequently, is a conservative selection in that it prioritises sample fidelity ahead of completeness. \citet{CidFernandes2010} found that only forty per cent of emission line galaxies in the "right wing" and AGN areas of the BPT diagram may be detected with $\text{S/N} >3 $ in both H$\beta$ and [\ion{O}{iii}]~$\lambda 5007$, and thus be selected as AGN. That is to say, in order to preserve sample fidelity, the criteria of \citet{Kewley2001} may miss a majority of narrow-line AGN.

To account for this, \citet{CidFernandes2010} proposed selecting narrow-line AGN on the basis of the two stronger emission lines, H$\alpha$ and [\ion{N}{ii}]~$\lambda 6583$. Using these two emission lines, AGN may be selected by comparing the ratio of [\ion{N}{ii}]~$\lambda 6583$ to t H$\alpha$, to the equivalent width of H$\alpha$. The resultant so called `WHAN' diagram \citep{CidFernandes2011} categorises emission line galaxies (ELGs) as having $\text{EW}_{\text{H}\alpha} > 3\,\text{\AA}$, and:
\begin{itemize}
\item star-forming, $\log_{10} \bigg(\frac{\text{[\ion{N}{ii}]~$\lambda 6583$}}{\text{H}\alpha}\bigg) < -0.4$\\
\item weak AGN, $\log_{10} \bigg(\frac{\text{[\ion{N}{ii}]~$\lambda 6583$}}{\text{H}\alpha}\bigg) > -0.4$
and $\text{EW}_{\text{H}_\alpha} < 6\,\text{\AA}$\\
\item strong AGN, $\log_{10} \bigg(\frac{\text{[\ion{N}{ii}]~$\lambda 6583$}}{\text{H}\alpha}\bigg) > -0.4$ and $\text{EW}_{\text{H}_\alpha} > 6\,\text{\AA}$\\
\end{itemize}
where the category `weak AGN', are the less powerful counterparts to the `strong AGN', but where the AGN is still considered to be the dominant ionisation mechanism \citep{CidFernandes2011}. Galaxies with $\text{EW}_{\text{H}\alpha} < 3\,\text{\AA}$ are considered to be passive in nature.

To select our AGN using the WHAN criteria, we select the best spectrum available for galaxies in our group sample with $nQ\geq3$, i.e., a greater than $90\,$per cent confidence in the measured redshift \citep{Driver2011, Liske2015}. We use only spectra with H$\alpha$ and [\ion{N}{ii}]~$\lambda 6583$ in emission. This requires that the Gaussian has been successfully fit to each line in {\sc SpecLineSFRv05}, the values and errors on flux and equivalent widths to be greater than zero, and the signal to noise ratio on both lines is greater than 3. For AAT obtained spectra we use only those spectra unaffected by fringing \footnote{Approximately $5\,$per cent of AAOmega spectra suffer from a time-dependent fringing artefact \citep{Hopkins2013}; this is flagged in the GAMA catalogue {\sc AATSpecAllv27}.}. Furthermore, it is difficult to obtain reliable mass estimates for broad-line AGN \citep{Gordon2017} and we hence exclude them from this analysis. Thus, we only use galaxies that do not have a broad component to the H$\alpha$ line, i.e., a single Gaussian fit is preferred.

\begin{figure}
\includegraphics[width=\columnwidth]{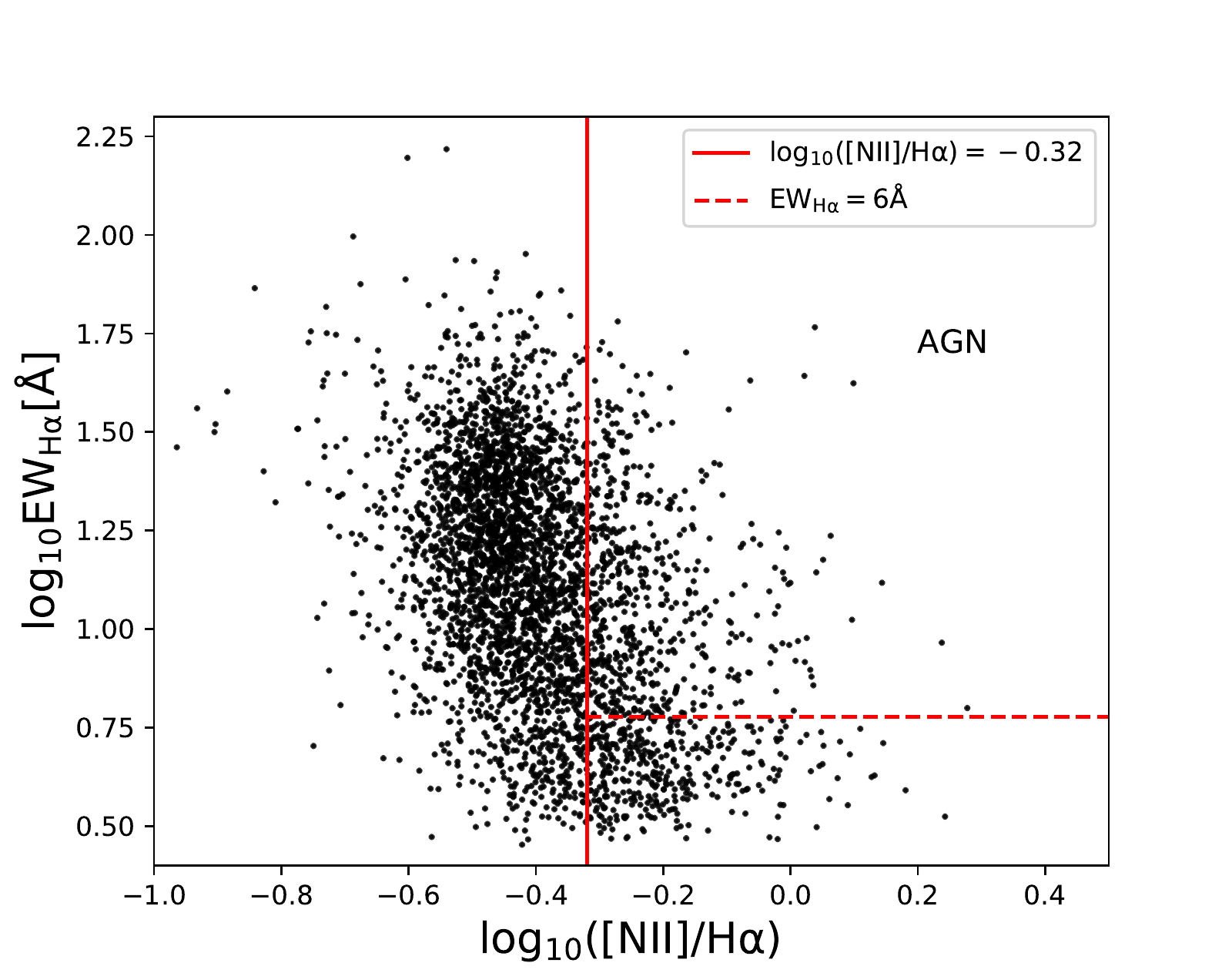}
\caption{A WHAN diagram of the emission line galaxies selected in our group sample. The horizontal red dashed line is used to segregate AGN from weak AGN composites, and LIERs. The vertical red line is $\log_{10}(\text{[\ion{N}{ii}]$~\lambda 6583$}/\text{H}\alpha) = -0.32$ which defines the boundary between the star-forming and AGN regions of the plot. To be classified as an AGN a galaxy must be located above the red dashed line and to the right of the red solid line. Note the absence of galaxies with $\log_{10}(\text{EW}_{\text{H}\alpha}/\text{\AA})<0.5$, a consequence of the stringent selection criteria applied.}
\label{WHAN}
\end{figure}

These criteria select 2864 emission line galaxies from the 7498 in our group sample. In order to account for Balmer absorption, the flux and equivalent width of the H$\alpha$ line are corrected, as per \citet{Hopkins2013}, by:
\begin{equation}
S_{\text{H}\alpha\text{, intrinsic}} = S_{\text{H}\alpha \text{, observed}} \Bigg(\frac{\text{EW}_{\text{H}\alpha} + 2.5\,\text{\AA}}{\text{EW}_{\text{H}\alpha}}\Bigg)
\label{EQfluxcorrect}
\end{equation}
The WHAN diagram of the emission line galaxies is shown in Figure \ref{WHAN}. Note, that because of the strict criteria imposed to ensure the reliability of the detected emission lines, the passive galaxy section of this diagram is barren.

\subsubsection{Sample contamination}
\begin{figure}
\includegraphics[width=1.1\columnwidth]{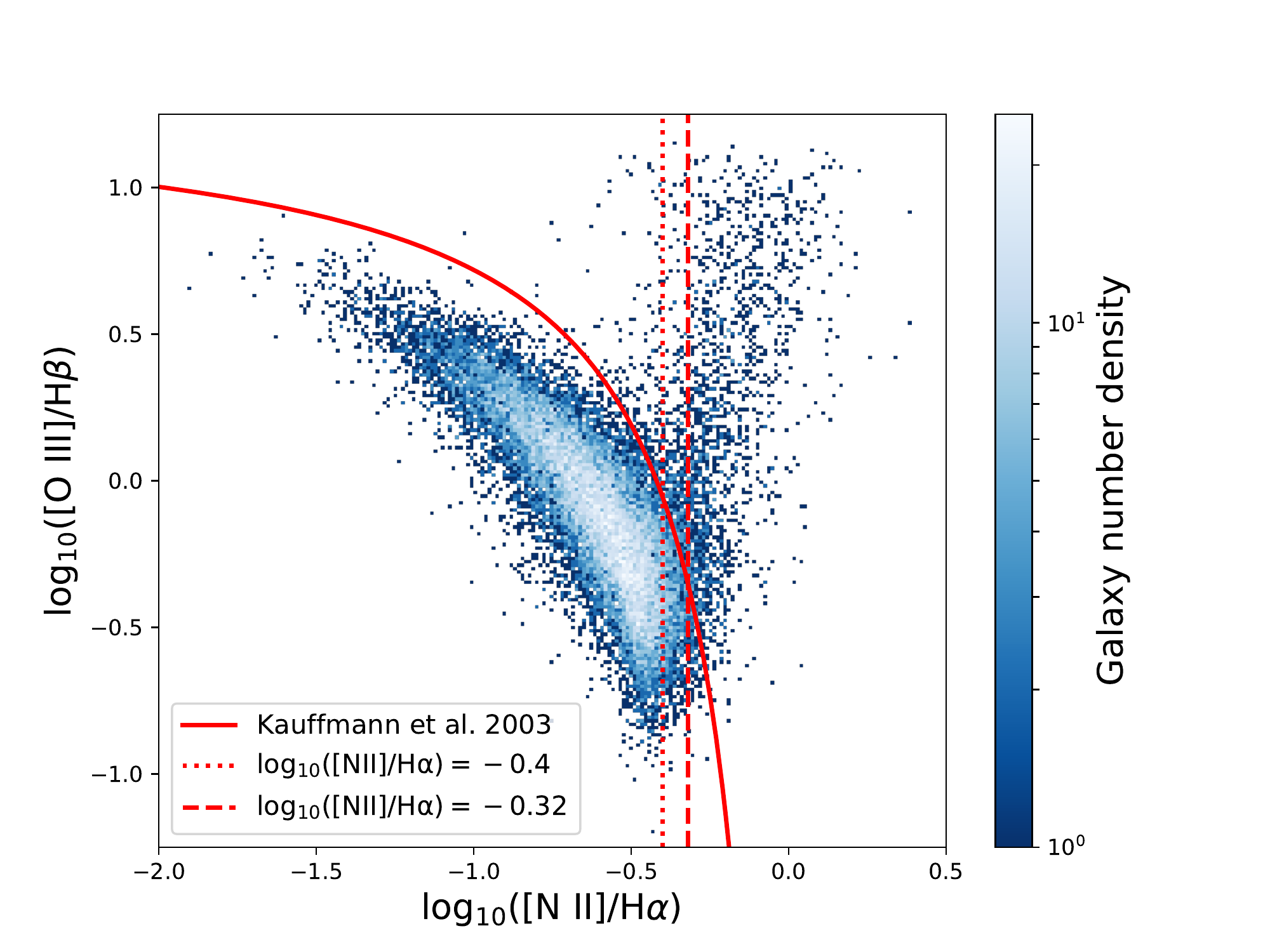}
\caption{A BPT diagram of the emission line galaxies in GAMA. The \citet{Kauffmann2003} line is plotted as the soid red curve. The red dotted line shows $\log_{10}(\text{[\ion{N}{ii}]~$\lambda6583$}/\text{H}\alpha) = -0.4$, often used to select AGN in WHAN diagnostic criteria. The red dashed line shows $\log_{10}(\text{[\ion{N}{ii}]~$\lambda6583$}/\text{H}\alpha) = -0.32$ used to select AGN in this work. The large fraction of star forming galaxies located between the two vertical lines is readily apparent.}
\label{BPT}
\end{figure}

Although such an AGN selection will be more complete than more conservative methods, the level of contamination from galaxies ionised by a non-AGN source will naturally be higher. Given that the WHAN and BPT diagrams share an axis, this contamination can be visualised by over-plotting the \citet{CidFernandes2010} AGN/SF segregation on to the BPT (See Figure \ref{BPT}).
To quantify, and minimise this, one can compare the classifications on both the WHAN and BPT of galaxies where all four BPT lines are detected with $\text{S/N}>3$. For this, we classify those galaxies satisfying the \citet{Kewley2001} criteria (equation \ref{EQKe01}) as AGN, those satisfying the \citet{Kauffmann2003} criteria,
\begin{equation}
\log_{10}\Bigg(\frac{\text{[\ion{O}{iii}]~$\lambda5007$}}{\text{H}\beta}\Bigg) <\frac{0.61}{\log_{10}\Big(\frac{\text{[\ion{N}{ii}]~$6583$}}{\text{H}\alpha}\Big)-0.05} + 1.3
\label{EQKa03}
\end{equation}
as star-forming galaxies, and those galaxies not satisfying either criteria as composite sources. 

Applying these criteria to WHAN selected AGN, shows that contamination of the AGN population by star-formers is limited to those AGN with the lowest [\ion{N}{ii}]~$\lambda6583$:H$\alpha$ ratios. The [\ion{N}{ii}]~$\lambda6583$:H$\alpha$ used in the WHAN diagram is derived from the \citet{Stasinska2006} AGN/star-forming segregation. Transposing the more conservative \citet{Kauffmann2003} AGN/star-forming segregation on to the WHAN diagram would result in galaxies with $\log_{10}(\text{[\ion{N}{ii}]~$\lambda6583$/H}\alpha) < -0.32$ being classified as star-forming. Indeed, when only galaxies with $-0.4 < \log_{10}(\text{[\ion{N}{ii}]~$\lambda6583$/H}\alpha) < -0.32$ are considered, star-formers make up $75.88_{-1.13}^{+1.06}$ per cent of the population with all four BPT lines at $\text{S/N} >3$. When only galaxies with $\log_{10}(\text{[\ion{N}{ii}]~$\lambda6583$/H}\alpha) > -0.32$ are considered, the contamination of the sample by star forming galaxies is reduced to $11.07_{-0.85}^{+0.99}$ per cent. Given the decreasing detections of all four BPT lines with increasing [\ion{N}{ii}]~$\lambda6583$:H$\alpha$ ratio shown in \citet{CidFernandes2010}, we take this as an upper limit on the star-forming contamination of the ELG sample.

While altering the minimum [\ion{N}{ii}]~$\lambda6583$:H$\alpha$ ratio of our sample will reduce the contamination from pure star-forming sources, the contribution from the composite population is substantial. There are various arguments as to the nature of this population with a variety of sources thought to contribute. \citet{Kewley2001} argue that galaxies can be classified as a composite solely as the result of ionisation from an extreme starburst phase. However, the sample contamination from such sources will be minimal. Taking the \citet{Poggianti2000} e(b) classification as a proxy for the starburst population, we find these galaxies constitute $<1$ per cent of the ELG population in GAMA. Ionisation from post-AGB stars is likely to be a considerable source of contamination for the composite population. Furthermore, the \citet{Kewley2001} AGN population will be contaminated by low-ionisation emission regions (LIERs). Both of these contaminants present with weak hydrogen lines \citep[Fig. 2 of][]{Marshall2017, Sanchez2017}, and thus their contribution to an AGN selection can be minimised by only selecting AGN with strong H$\alpha$, i.e., $\text{EW}>6\,\text{\AA}$, emission.

To achieve a balance between high sample completeness and fidelity, we select our AGN using a conservative variant of the WHAN diagnostic criteria. Specifically we classify as AGN galaxies with $\log_{10}(\text{[\ion{N}{ii}]~$\lambda6583$/H}\alpha) > -0.32$ and $\text{EW}_{\text{H}\alpha}>6\,\text{\AA}$. This produces a sample of 451 AGN in our selected groups.

\section{Observations and Analysis}
\label{Sobs}
\subsection{The group AGN fraction}
\begin{figure}
\includegraphics[width=1.1\columnwidth]{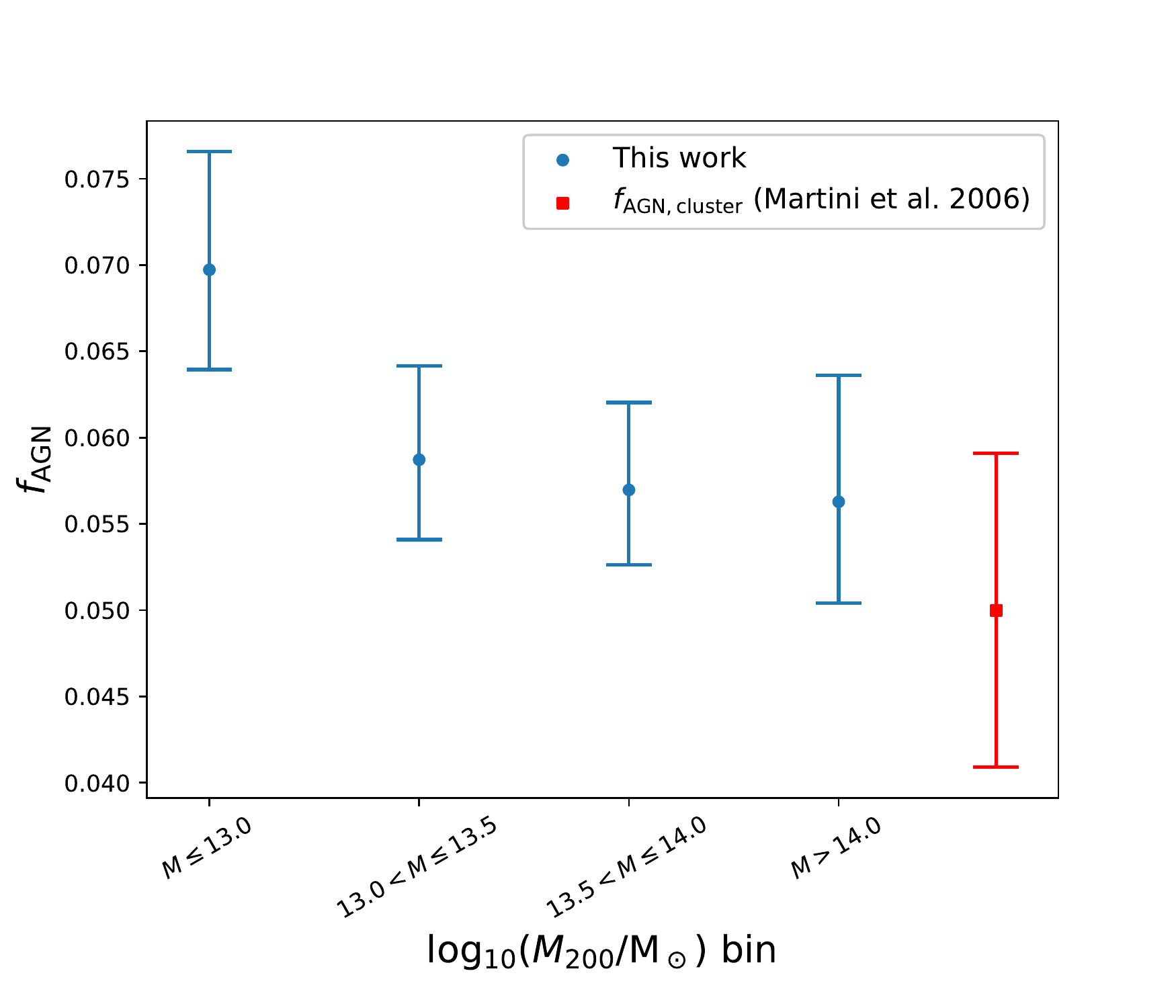}
\caption{The fraction of the galaxy population hosting an AGN in bins of group mass, $M_{200}$. The AGN fraction is flat for $M_{200}>10^{13.0}\,\text{M}_\odot$, and marginally elevated at lower halo masses. The red square shows the AGN fraction in clusters, and 1$\sigma$ errors, obtained from \citet{Martini2006} for comparison.}
\label{AGNfracNfof}
\end{figure}

Taking the group population as a whole, we find that AGN constitute $6.01_{-0.26}^{+0.29}\,$per cent of the galaxies more massive than $10^{9.9}\,\text{M}_\odot$ in our sample. Given that the masses of our groups span three orders of magnitude (See Table \ref{Tgroup}) we wish to determine whether this AGN fraction is dependent on the mass of the group. In Figure \ref{AGNfracNfof} we split our group sample in to four bins of $M_{200}$. We show that the AGN fraction is approximately flat at a halo mass of  $\log_{10}(M_{200}/\text{M}_\odot) > 13$. Lower mass groups have a higher, if not significantly so, AGN fraction than higher mass groups. For all fractional calculations, both here and throughout this paper, the errors are binomial as per \citet{Cameron2011}.

\subsection{AGN in group projected phase-space}
It has been shown that for massive clusters AGN, are preferentially found in the infall regions of those clusters, and are less likely to be found in the cluster centre \citep[][]{Haines2012, Pimbblet2013}. To test if this is the case in lower mass groups, we stack our selected groups into one large `super-group' of 7498 galaxies. We then analyse the positions of those galaxies in the projected radius versus velocity difference phase-space plane. In order for this projected phase-space to be useable for our stacked group we must normalise the dimensions. For the projected separation, we use the projected separation of a galaxy from the iterative centre of the group, and normalise by $R_{200}$. To calculate the velocity difference, we use the difference between the redshift of the galaxy and the median redshift of the group, i.e., $\Delta v = c(z_\text{gal}-z_\text{group})/(1+z_\text{group})$. This is normalised by the group velocity dispersion.

\begin{figure*}
	\centering
	\begin{subfigure}{\columnwidth}
	\includegraphics[width=\columnwidth]{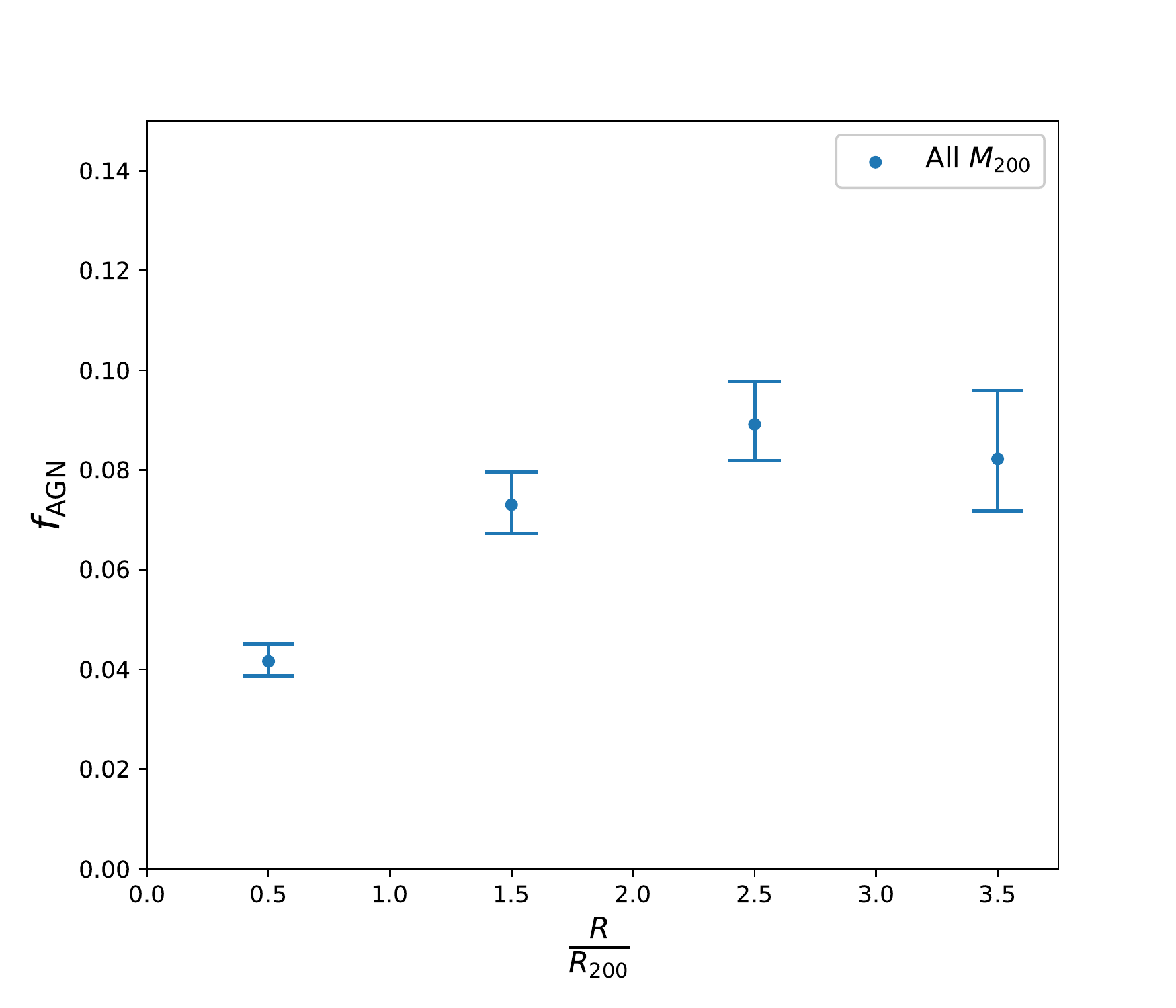}
	\caption[]{}
	\end{subfigure}
	\begin{subfigure}{\columnwidth}
	\includegraphics[width=\columnwidth]{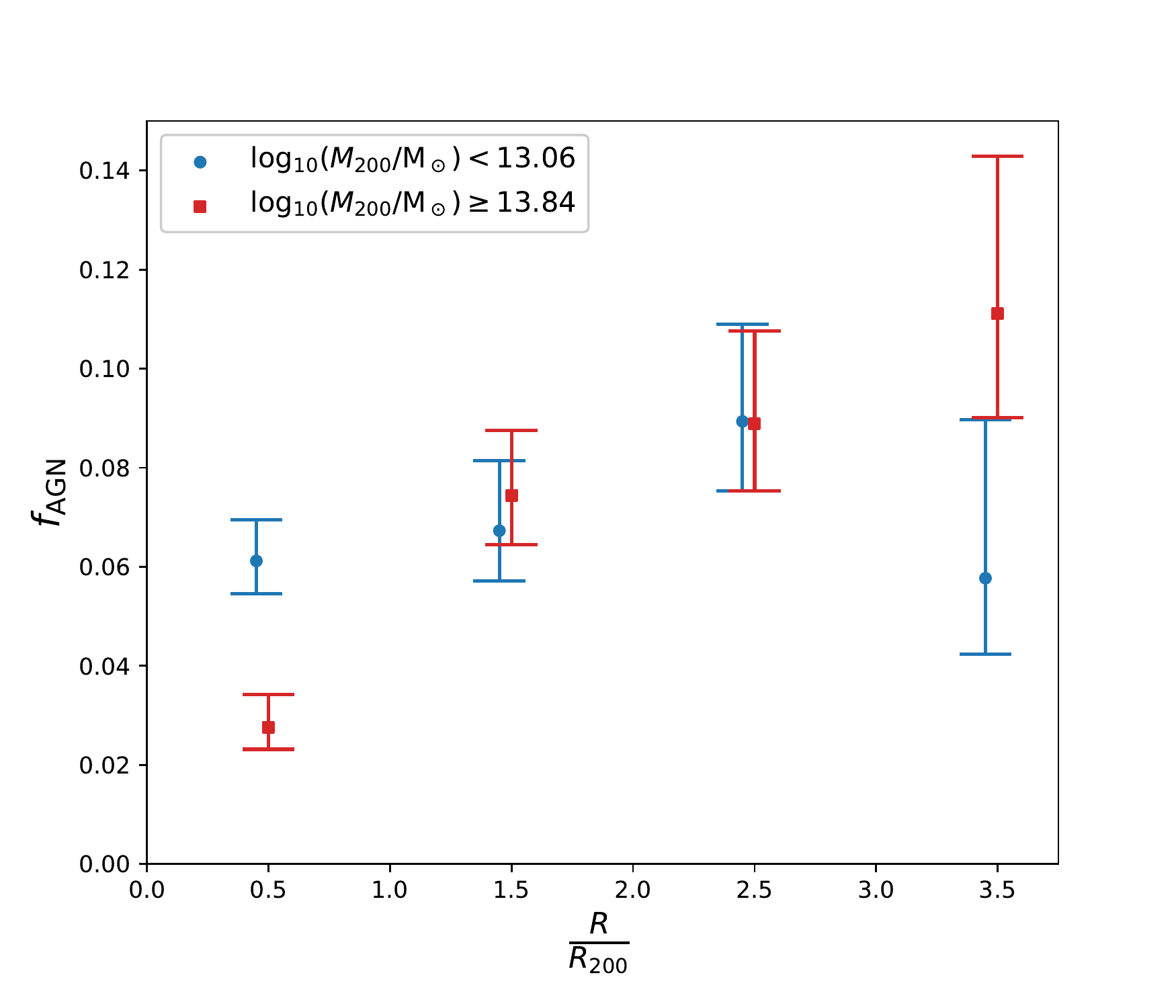}
	\caption[]{}
	\end{subfigure}
\caption{The effect of projected separation from group centre on AGN fraction. In panel (a) the entire group sample is considered. Panel (b) splits the sample into the upper and lower quartiles of the group mass distribution, showing that low mass groups don't have the deficit of centrally located AGN seen in high-mass groups and clusters \citep{Ehlert2013, Pimbblet2013}. For panel (b) blue circles represent groups with $\log_{10}(M_{200}/\text{M}_\odot) < 13.06$, while groups with $\log_{10}(M_{200}/\text{M}_\odot) > 13.84$ are shown by red squares, these are marginally offset along the $x$-axis for clarity. For both panels the bin width is $R_{200}$ for all bins.}
\label{phasespaceR}
\end{figure*}

\citet{Ehlert2013} and \citet{Pimbblet2013} showed that AGN fraction increases with projected radius from the cluster centre. Given this established result for clusters, we test whether the same result holds for galaxy groups. In Figure \ref{phasespaceR} we show the same trend for our group sample. In order to test whether this effect is biased by the large mass range of groups in our sample, we repeat this test for the upper and lower quartiles of the mass distribution of our group sample. This shows that the low AGN fraction inside $R_{200}$ is driven by high-mass groups, where the AGN fraction is lower at $R<R_{200}$ at $2.9\sigma$ confidence.

\begin{figure*}
	\centering
	\begin{subfigure}{\columnwidth}
	\includegraphics[width=\columnwidth]{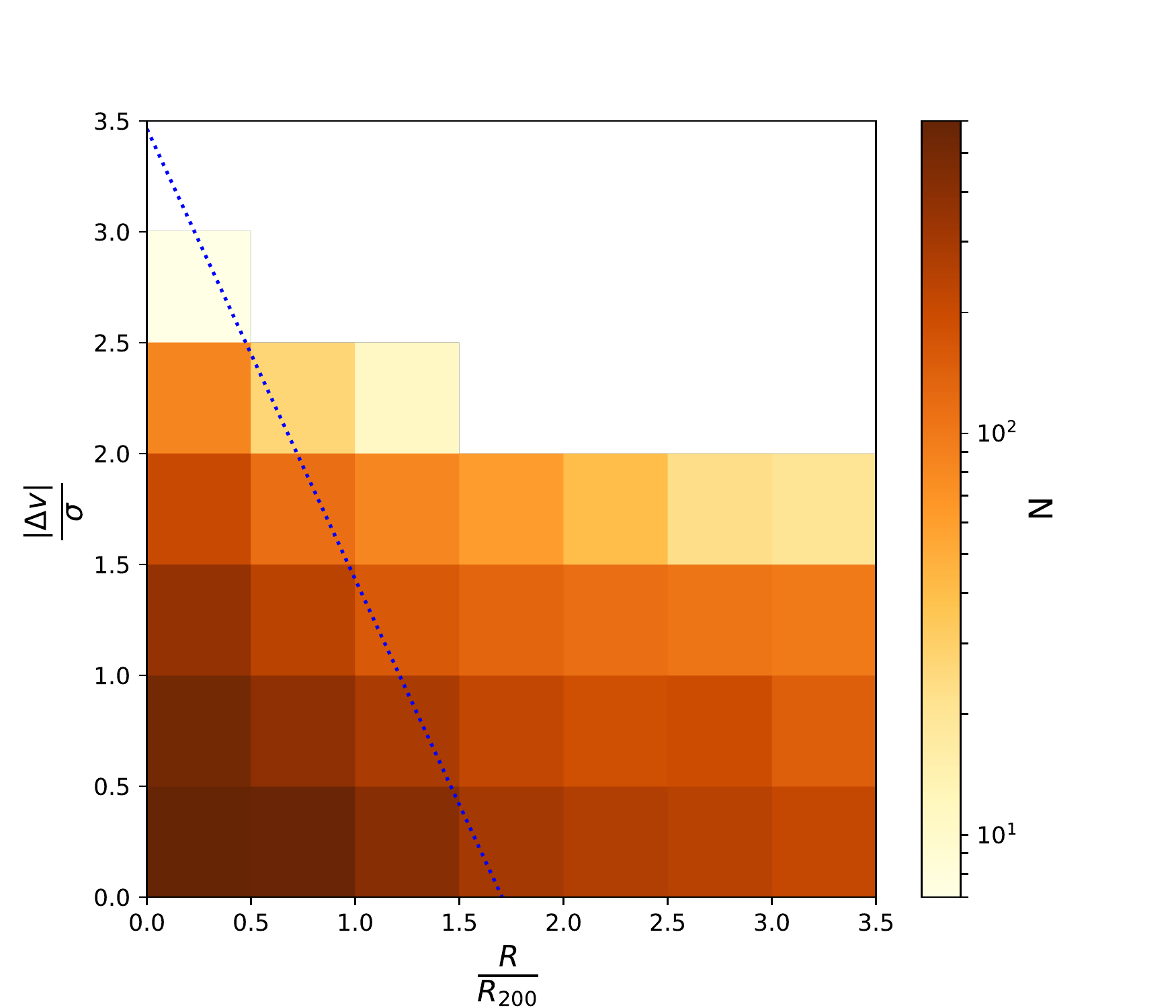}
	\caption[]{All galaxies with $\log_{10}(M_{*}/\text{M}_\odot)>9.9$}
	\end{subfigure}
	\begin{subfigure}{\columnwidth}
	\includegraphics[width=\columnwidth]{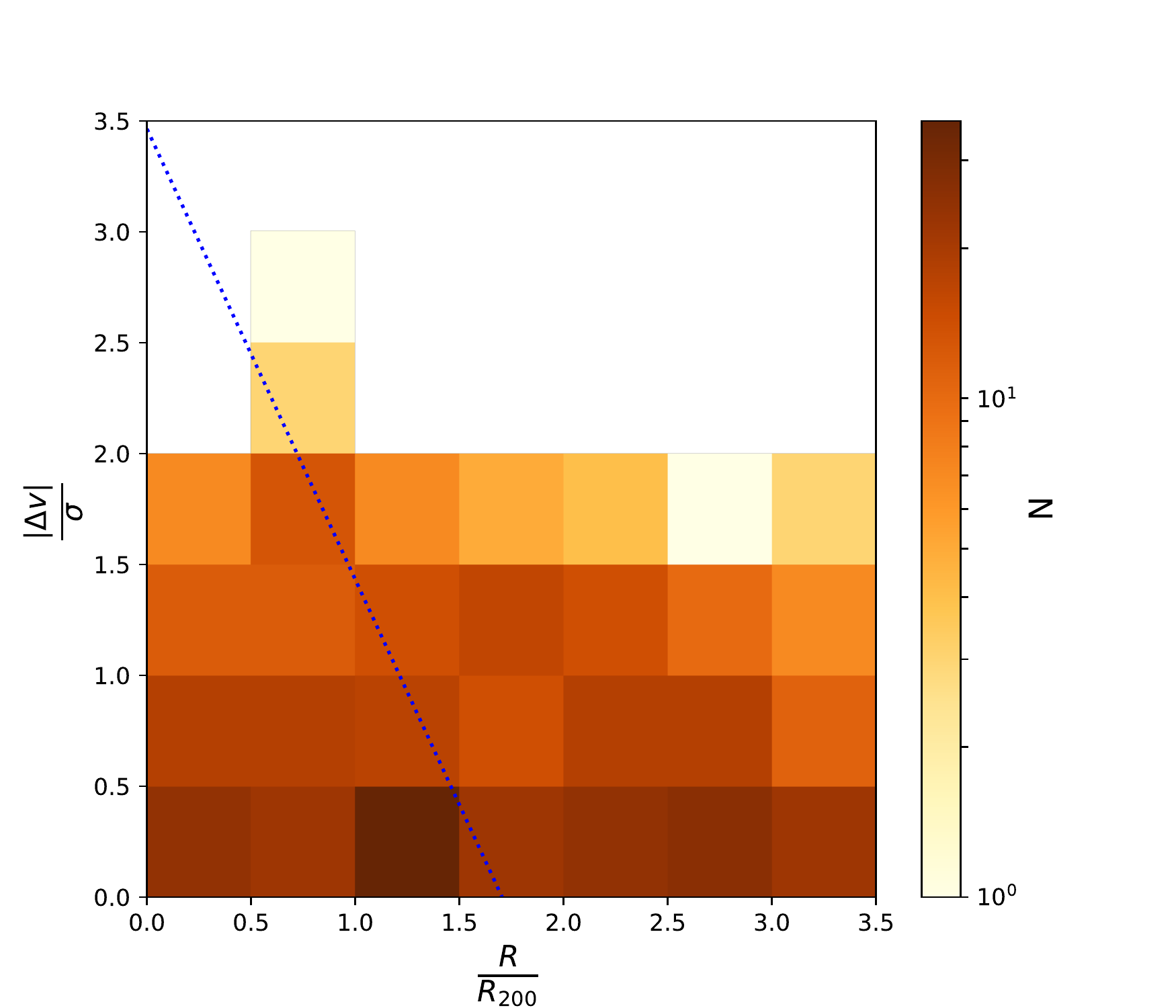}
	\caption[]{AGN}
	\end{subfigure}
	\begin{subfigure}{\columnwidth}
	\includegraphics[width=\columnwidth]{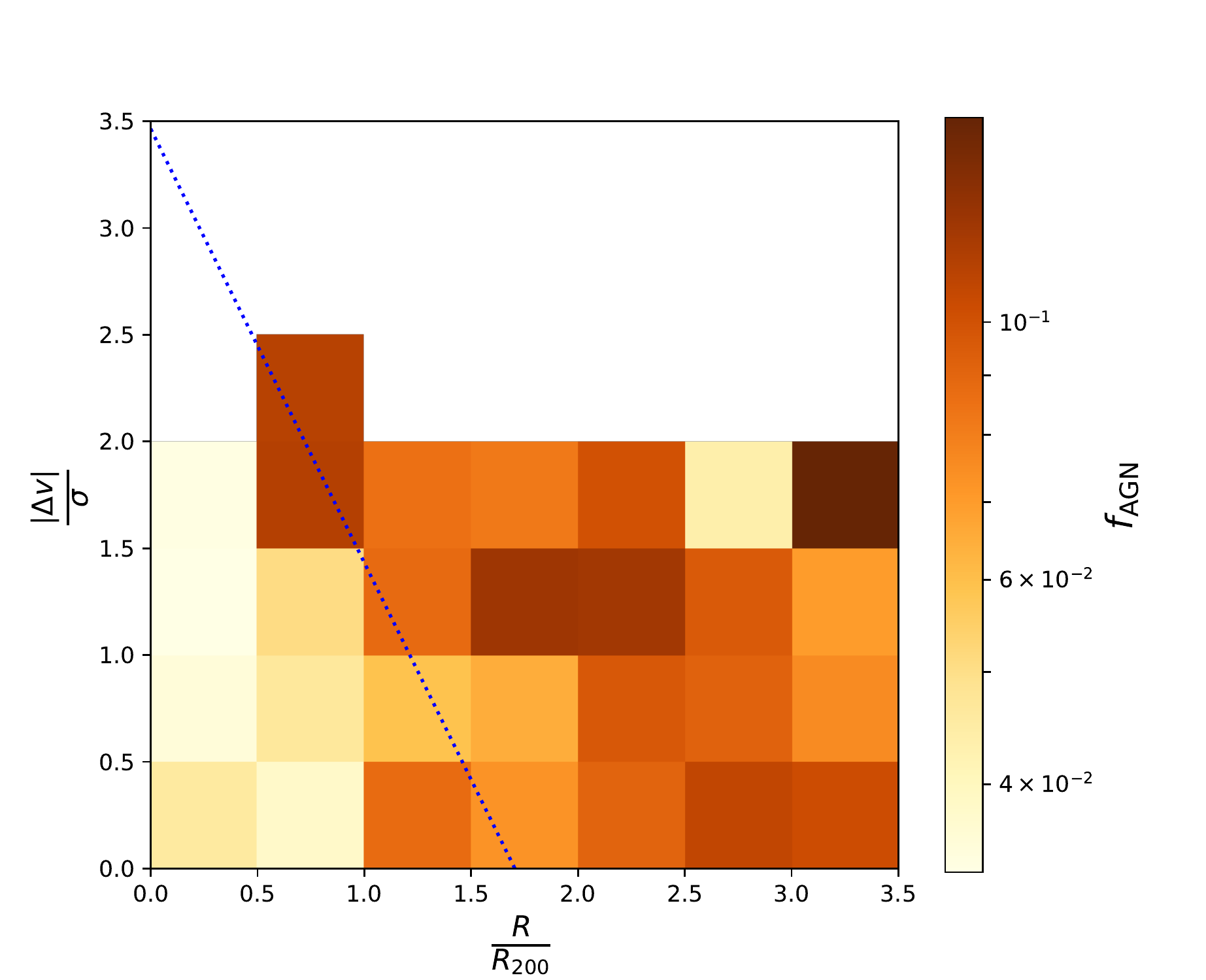}
	\caption[]{AGN fraction}
	\end{subfigure}
\caption{Two dimensional histograms showing the projected phase-space distributions of our stacked group of 7498 galaxies. Panel (a) shows the number density of all galaxies, panel (b) shows the number density of AGN, and panel (c) shows the AGN fraction. The blue dotted line shows the modelled infall curve of \citet{Oman2013}. The median absolute error for the AGN fractions in panel (c) is 0.02.}
\label{phasespace2d}
\end{figure*}

In Figure \ref{phasespace2d} we show the relative number densities of galaxies and AGN in projected phase-space and the expected infall curve from the work of \citet{Oman2013}. This curve is defined as
\begin{equation}
\frac{|v|}{\sigma_{\text{group, 3D}}} = -\frac{4}{3}\frac{R}{R_{\text{virial}}} +2
\label{OmanEQ}
\end{equation}
where $R_{\text{viral}} =\frac{2.5}{2.2}R_{200}$ \citep{Oman2013}, and $\sigma_{\text{group, 3D}}$ is the three-dimensional equivalent of $\sigma_{\text{group}}$, i.e., $\sqrt{3}\sigma_{\text{group}}$ \citep{Barsanti2017}.
By comparing with this expected infall curve, Figure \ref{phasespace2d} shows qualitatively that, as one might expect for virialised structures, the number density of galaxies is higher below this curve. The number density of AGN, however, is seen to be more widely distributed, and the AGN fraction is highest in the infall regions, i.e. above the \citet{Oman2013} curve.

The AGN fractions calculated for the bins in Figure \ref{phasespace2d} have a median absolute error of 0.02. To reduce this error and better quantify the difference in AGN triggering likelihood with environment, we calculate the AGN fractions for the virialised and infalling galaxy populations as defined by equation \ref{OmanEQ}.
For our stacked group, the AGN fraction in the virialised region is $4.50_{-0.32}^{+0.36}$ per cent compared to $7.56_{-0.41}^{+0.46}$ per cent in the infall regions. Across our sample, AGN are morel likely to be found in infalling galaxies with $3.9\sigma$ confidence.

\begin{figure*}
	\centering
	\begin{subfigure}{\columnwidth}
	\includegraphics[width=0.95\columnwidth]{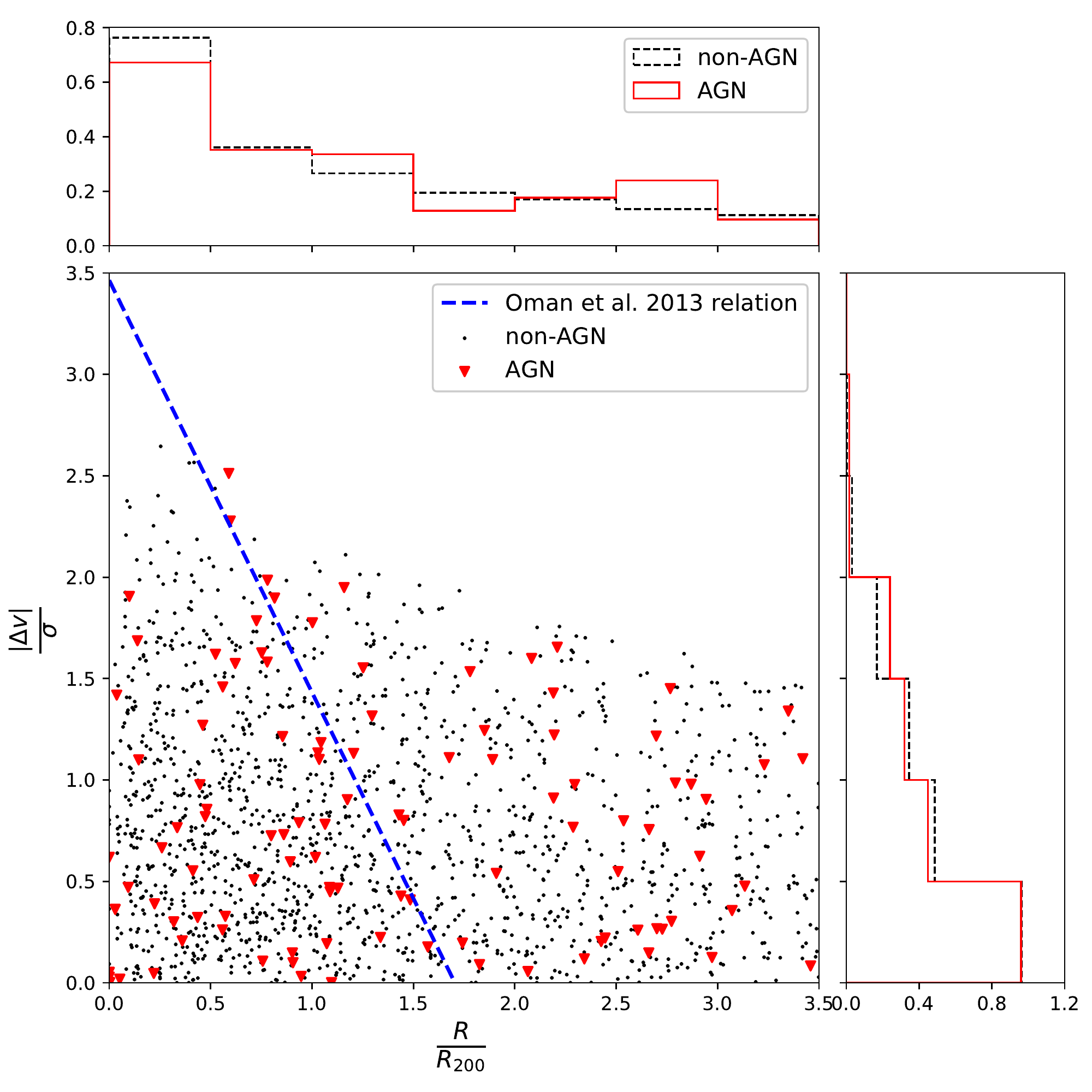}
	\caption[]{$\log_{10}(M_{200}/\text{M}_\odot)< 13.06$}
	\end{subfigure}
	\begin{subfigure}{\columnwidth}
	\includegraphics[width=0.95\columnwidth]{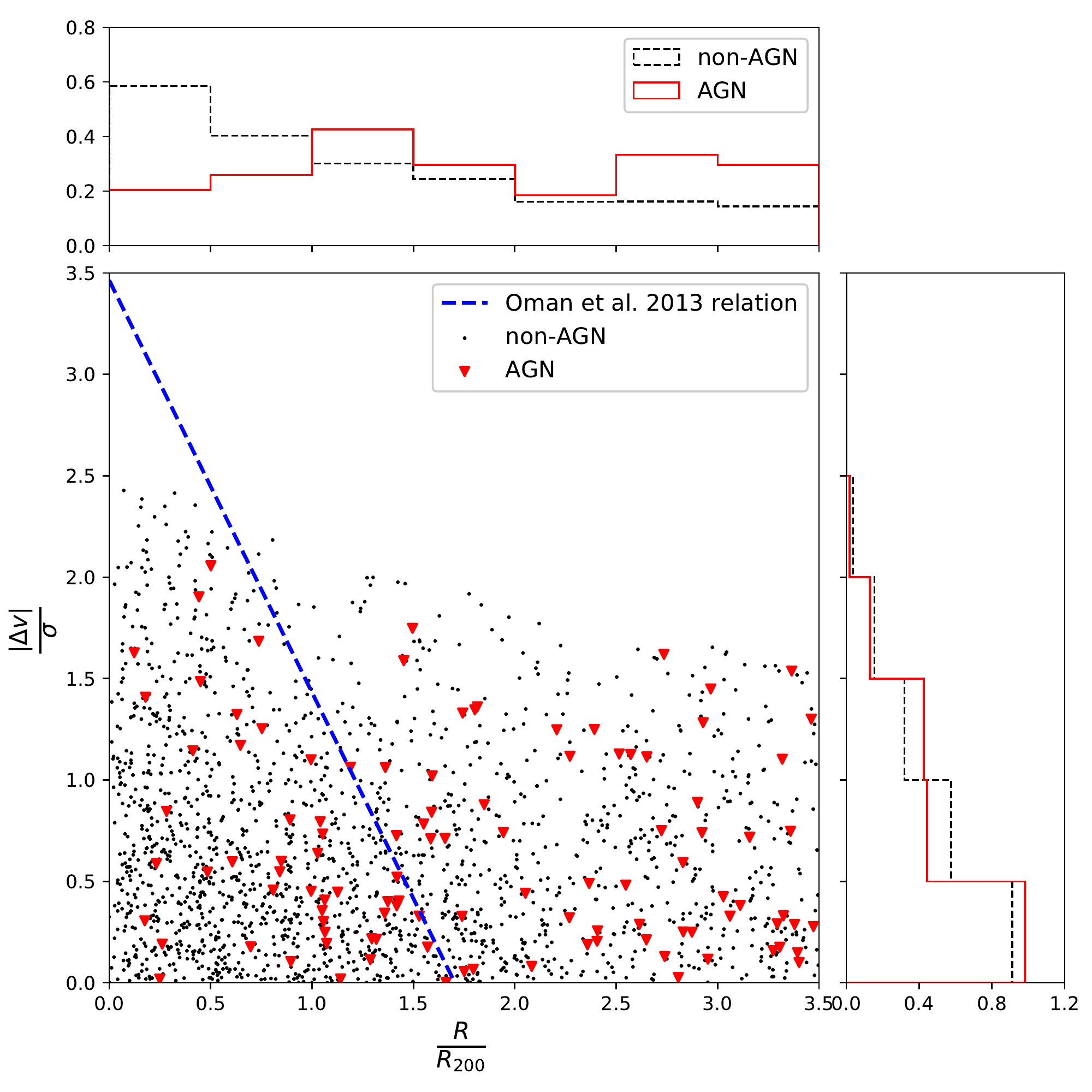}
	\caption[]{$\log_{10}(M_{200}/\text{M}_\odot)> 13.84$}
	\end{subfigure}
\caption{The projected phase-space diagrams and histograms for groups in the lower (panel a) and upper (panel b) quartiles of the group halo mass distribution. Inactive galaxies are represented by black dots in the phase-space diagram and black dashed line in the histograms, and AGN are represented by red triangles and red solid lines in these plots respectively. The blue dashed line marks the \citet{Oman2013} infall curve. Panel b clearly shows a low number of AGN at low projected radii relative to the inactive population.}
\label{MbinnedPspace}
\end{figure*}

\begin{figure}
\includegraphics[width=\columnwidth]{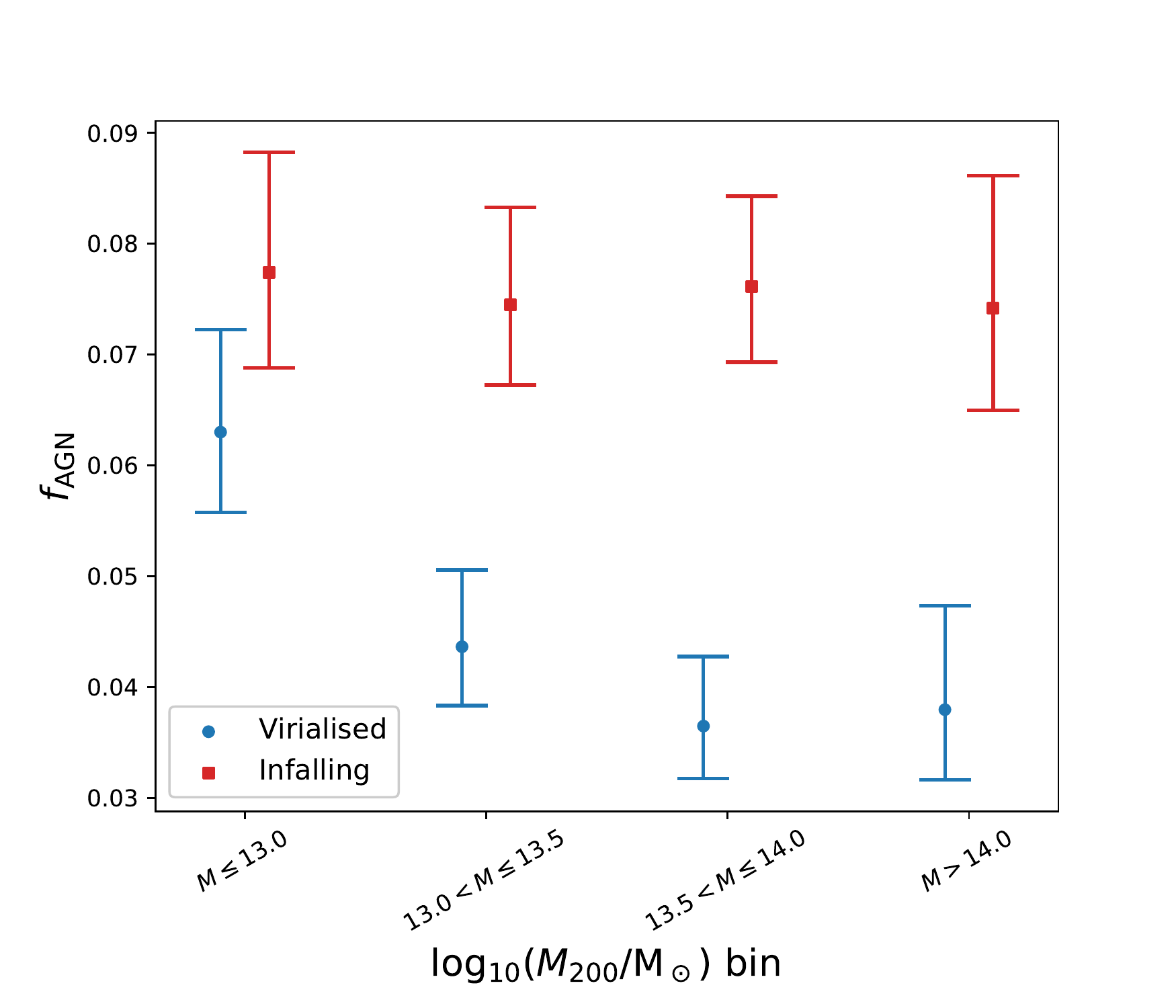}
\caption{A comparison of the AGN fractions of the virialised and infalling galaxy populations defined by equation \ref{OmanEQ} \citep{Oman2013}. Blue circles show the virialised AGN fraction, which decreases with halo mass for $\log_{10}(M_{200}/\text{M}_\odot)<13.5$. The infalling AGN fraction is represented by red squares and is flat across all halo masses in our sample. As with Figure \ref{phasespaceR} a marginal $x$-axis offset is used for the two populations to improve clarity.}
\label{MbinnedAGNfrac}
\end{figure}

As with the overall group AGN fraction, and the effect of projected separation from the group centre, we wish to test whether this preference for AGN to be found in the infalling population is driven by groups in a particular mass regime.
In Figure \ref{MbinnedPspace}, we show the galaxies and AGN in projected phase-space for the lower and upper quartiles, in terms of halo mass, of our sample. We also show the histograms for active and inactive galaxies in each of the projected phase-space dimensions, which clearly show a difference in the distribution of AGN at these halo mass regimes.
To quantify this, we attempt to constrain at what group mass the environmental transition occurs. In Figure \ref{MbinnedAGNfrac}, we bin the mass range of our sample and compare the likelihood of virialised and infalling galaxies hosting an AGN. This shows a decreasing AGN fraction amongst virialised galaxies with increasing group mass. At $\log_{10}(M_{200}/\text{M}_\odot) > 13.50$ the  observed AGN deficit in the group centre becomes significant at the $3.6\sigma$ level. For group halo masses greater than $10^{13.5}\,\text{M}_\odot$, $3.70_{-0.40}^{+0.50}\,$per cent of the virialised galaxy population host an AGN, in contrast to $7.55_{-0.57}^{+0.66}\,$per cent of the infalling population.

\section{Discussion}
\label{Sdiscuss}
\subsection{AGN in large-scale structure}
The dynamics of large-scale structure (LSS) change with structure size. For instance, one would not expect that a galaxy pair would have a similar kinematic environment to a larger group of many tens of members, the latter of which may well be virialised. Galaxy pairs, especially a close pair, and small groups of only a few members are likely to be directly interacting \citep{Hickson1997, Robotham2014}, a widely considered trigger for nuclear activity within galaxies \cite[e.g.,][]{Sanders1988, Krongold2002, Ellison2011}. This is supported by observations that both pairs and compact groups exhibit higher nuclear activity compared to field environments, with AGN fractions approaching $\sim20\,$per cent in the closest pairs \citep{Rubin1991, Menon1995, Ellison2011}.

Pairs and compact groups represent the smallest super-galactic structures in the universe. At the other end of the scale are massive clusters. In contrast to to smaller structures, clusters have AGN fractions of just $\sim5\,$per cent \citep{Martini2006, Arnold2009}. This is lower than the field AGN fraction of $8\pm0.13\,$per cent\footnote{We have extrapolated the uncertainty on this value from Table 1 of \citet{Woods2007}.} \citep{Woods2007}. This could suggest that the environmental changes associated with moving up the LSS mass function first act to increase AGN triggering opportunities through direct interaction, before inhibiting AGN triggering in high mass structures.

As a mid-point in the LSS mass function, galaxy groups should therefore sit between small structures such as pairs and compact groups, and clusters in terms of AGN fraction. Indeed prior observations have shown galaxy groups to have AGN fractions in the range $\sim 7 - 9\,$per cent \citep{Shen2007, Arnold2009, Oh2014, Tzanavaris2014}. Our own observations are broadly consistent with this picture. We observe an AGN fraction of $6.01_{-0.26}^{+0.29}\,$per cent, although we note that this covers the mass range of our group sample, i.e., $11.53 \leq \log_{10}(M_{200}/\text{M}_\odot)\leq 14.56$. Consequently this may be driven by the high mass end of our group sample. Taking only the lowest quartile of the group mass range, the AGN fraction is $6.68_{-0.53}^{+0.62}\,$per cent. This is perfectly consistent with the overall value observed in our sample, but is closer to the expected value from previous studies.

\subsection{Galaxy location within the group structure}
As well as the galactic environment changing with group mass, the environmental conditions are a function of location within the group. 
Extending on the work of \citet{Haines2012} and \citet{Pimbblet2013}, we have shown in Figure \ref{phasespaceR} that, although the radial AGN fraction of our groups decreases inside $R_{200}$, this effect is driven by high-mass groups. For low-mass groups, we show the AGN fraction out to $3.5R_{200}$ to be approximately flat. This suggests that the central regions of low-mass groups are similar to the more extended regions of those groups in terms of their ability to host AGN.

To conduct a more physically motivated analysis, we use the criteria of \citet{Oman2013} to split the groups into infalling and virialised populations. This shows that AGN are significantly ($\sim3.9\sigma$) more likely to be found in infalling galaxies than virialised galaxies. As with the radial AGN fraction, this is primarily driven by the high mass groups in our sample. Figure \ref{MbinnedAGNfrac} shows the infall AGN fraction remains flat at $\sim7.5\,$per cent across the mass range of our group sample, comparable to the field AGN fraction \citep{Woods2007}. This is suggestive that the act of falling into a group does not increase the likelihood of a galaxy hosting an AGN.
The AGN fraction in the virialised region of the groups however, is shown to decrease with increasing halo mass. This effect becomes significant at the $3.6\sigma$ level for groups with $M_{200}>10^{13.5}\,\text{M}_\odot$, where only $\sim3.5\,$per cent of virialised galaxies host an AGN. That is to say, as one might expect, our high mass groups start to show the same observational properties with regard to AGN fostering that has been previously observed in clusters \citep{Gilmour2007, Gavazzi2011, Haines2012, Pimbblet2012, Pimbblet2013}.

The decreasing AGN fraction in virialised galaxies with increasing group mass leads to the inference that the physics of the group core evolves with group mass. Indeed this should be expected if groups are simply low-mass and less evolved analogues of clusters. In relaxed clusters, the cores are virialised and consequently the frequency of low-speed interactions that may act as a trigger for AGN is reduced. As a group evolves into a cluster one would expect the dominance of this effect to increase, and consequently this mechanism may well play a role in the low virialised AGN fraction we see in our high-mass groups.

A second mechanism that may contribute to the absence of AGN in the centres of large groups is the temperature of the IGM. In clusters, it has been suggested that the high temperature of the ICM prevents the matter from being accreted onto a galaxy and hence starves any potential AGN of a fuel supply \citep{Davis2016, Davies2017}. As one goes up the group mass function, the IGM temperature increases \citep{Shimizu2003}, and hence so does the effectiveness of this mechanism. Given that we see no substantial lack of AGN in the centres of low-mass groups, our observations may support such a model.

The inhibition of nuclear activity in the virialised region of phase-space of large groups may also be caused by RPS. The infall regions of clusters have been shown \citep{Marshall2017} to provide the appropriate gas pressures to compress and destabilise intra-galactic gas \citep{Schulz2001, Tonnesen2009}, potentially fuelling nuclear activity \citep{Marshall2017}. In the cores of clusters however, galaxies are subject to higher ram pressures that not only disrupt the internal galactic gas, but strip it from the galaxy \citep[e.g.,][]{Kenney2004}. Given that one would not expect the intense ram pressures required to strip a galaxy of gas in small groups, our observation of no AGN deficit in the central region of projected phase-space for low-mass groups is consistent with extreme RPS being a plausible mechanism for preventing nuclear activity in cluster cores.

The phase-space projections of galaxies within groups and clusters explain why the AGN fraction of the structure as a whole is lower than the AGN fraction in the field. If one considers only the infall region of the structure, the AGN fraction is shown to be a flat function of group halo mass (see red squares in Figuere \ref{MbinnedAGNfrac}). Combining the AGN fraction in the infall region for all our groups to reduce the uncertainty gives an AGN fraction of $\sim7.5\,$per cent, and is consistent with the AGN fraction in the field \citep{Woods2007}. However, if one then takes into account the core region of the cluster with its AGN deficit, this will act to reduce the overall AGN fraction. In low-mass groups where this effect is not observed, the infall AGN fraction is similar to the core AGN fraction, and thus similar to the field value.

\subsection{A comparison to radio AGN, star-forming galaxies, and the passive population}
The AGN used in this analysis have been selected optically, and hence their SMBH is accreting matter in a radiatively efficient manner. Such an accretion mode requires a cold gas fuel reservoir. In contrast, the SMBHs of radio selected AGN may inefficiently accrete matter whilst still powering the AGN. This inefficient accretion does not require a supply of cold gas, and the black hole may be `drip-fed' by either internal or external mechanisms \citep{Hardcastle2007, Best2012, Ellison2015}. Consequently the environmental effects inhibiting optical AGN in the cores of massive structures may not inhibit radio AGN. Indeed, the majority of low-powered radio AGN are found in the central galaxies in groups and clusters \citep{Best2007, Ching2017}. Therefore, these observations can only be used draw conclusions about the efficiently accreting AGN population.

Star forming galaxies, like optically selected AGN, require a fuel supply of cold gas. At the centres of massive groups, should that gas have been stripped on infall, and the IGM be too hot to accrete cold gas, then this will lead to the strangulation of star formation as well as reducing the AGN fraction in the group centre. Indeed, this effect is seen in massive clusters \citep{Lopes2017}. Additionally,  \citet{Barsanti2017} observe this effect with the fraction of star forming galaxies increasing in groups with $R/R_{200}$. Although their halo mass bins are larger than ours, their group sample shows a higher fraction of star forming galaxies at small projected radii than their cluster sample \citep[see Fig. 5 of][]{Barsanti2017}. This is consistent with what might be expected based on our demonstration of the importance of halo mass for galaxy nuclear activity. Furthermore, both \citet{Lopes2017} and \citet{Barsanti2017} observe the opposite effect for passive galaxies, with the passive population being more centrally located. As would be expected based on our observations, this effect is stronger in clusters than groups.

\subsection{Accounting for potential sources of bias}
\subsubsection{Stellar mass}
\begin{figure}
\includegraphics[width=1.1\columnwidth]{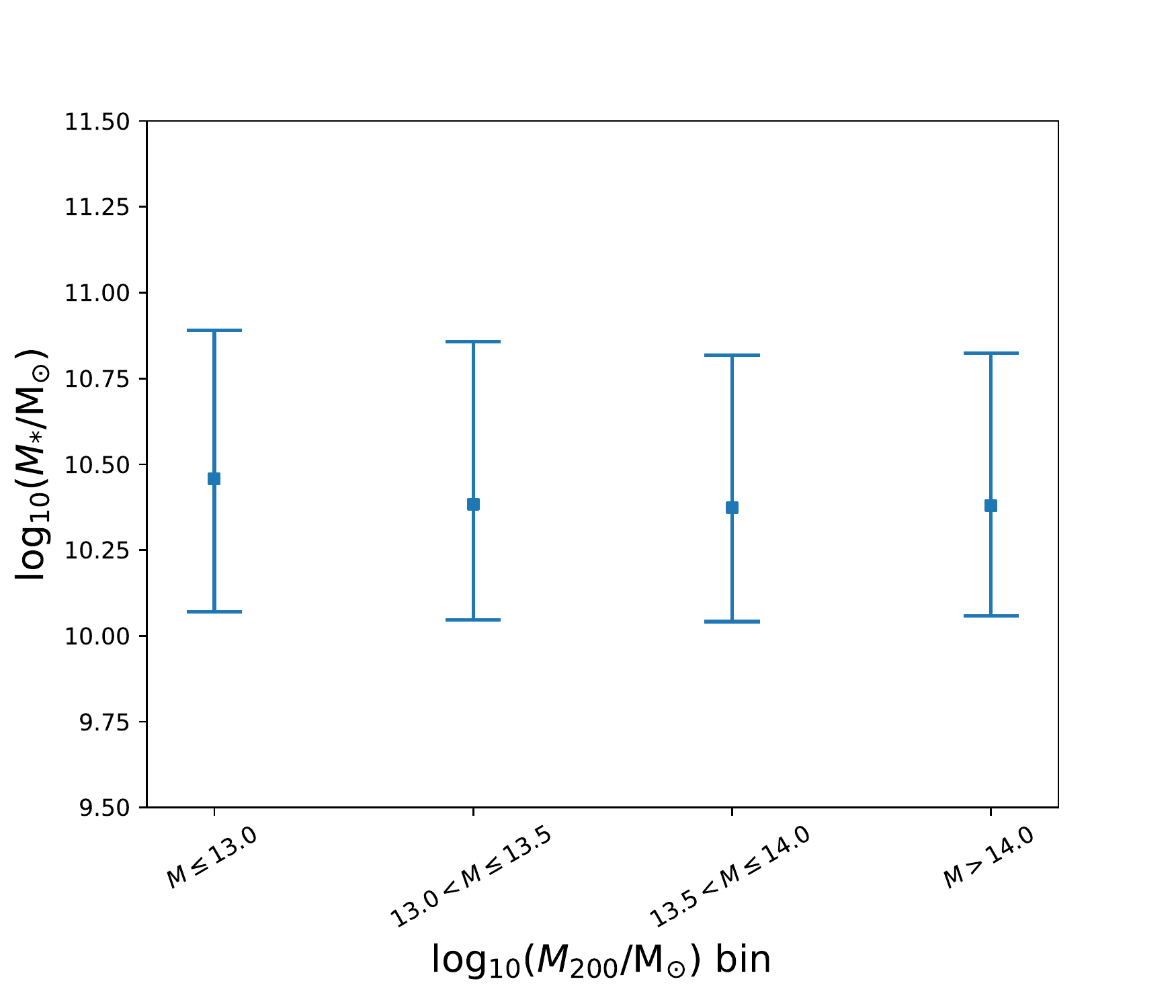}
\caption{The median stellar mass of galaxies with $\log_{10}(M_{*}/\text{M}_\odot)>9.9$ with group halo mass. Across the halo mass range of our sample there is no variation in the median galactic stellar mass, eliminating this as a potential source of bias for our observations. The error bars represent the 16th and 84th percentiles of the stellar mass distribution in each halo mass bin, and thus represent $68\,$per cent of the population.}
\label{m200mstar}
\end{figure}

The likelihood of a galaxy hosting an AGN increases with stellar mass \citep{Pimbblet2012, Pimbblet2013, Wang2017, Lopes2017}. Given that the central galaxies in groups are the most massive, the deficit of AGN in the cores of massive groups and clusters is unlikely to be a stellar mass bias. Moreover, if a dominant stellar mass bias were at play, one would expect to see the opposite observation. However, in low mass groups AGN are observed in the virialised region as frequently as in the infalling population. It is therefore necessary to confirm that the stellar mass distribution of galaxies within our group sample is not dependent in halo mass. In Figure \ref{m200mstar} we show the median stellar mass of galaxies within the same halo mass bins used in Figure \ref{MbinnedAGNfrac}, demonstrating there is no difference in the stellar mass distributions of our high and low mass groups.

\subsubsection{Group selection}
We have taken care throughout this work to ensure that bias from group selection effects is minimal in our results. Nonetheless, as with any observational work, one must consider whether such effects may drive a result. FoF algorithms are reliable but may under select from the extended group structure \citep{Barsanti2017}, whereas methods such as ours that select based on a radius and velocity are prone to contamination from field galaxies. That is to say, such methods prioritise completeness over fidelity \citep{Rines2005, Oman2016}.

In this work, to reduce the contamination of the group sample by interloping field galaxies, we exclude from selection galaxies that fall outside the NFW infall radius predicted for a group (see Section \ref{SSG3C}). However, by taking this approach there is every chance of excluding high-velocity infalling galaxies that genuinely are part of the group. To test whether this affected our results we reran the analysis on a group selection that did not account for the NFW profile. That is to say group membership was simply defined as $R<3.5R_{200}$; $|\Delta v| < 3.5\sigma_{\text{group}}$. This had no significant effect on our results with high mass groups having a central AGN deficit at $\sim4\sigma$ confidence.

To select our group members using the method described in Section \ref{S2groupmem}, the centre of the group was chosen to be the galaxy that was central galaxy in corresponding FoF group \citep[see Section 4.2 of][]{Robotham2011}. The logical alternative to this would be to centre the group selection on the brightest group galaxy (BGG). In $95\,$per cent of cases for groups with $N_{\text{FoF}}\geq5$, as we require for our group selection, the BGG was also the central galaxy in the group. Given this substantial overlap, we would expect altering the group selection in this way to have little effect on our results. To be certain of this, we reselected our groups using the BGG as the group centre. This resulted in a sample of $7482$ galaxies with $\log_{10}(M_*/\text{M}_\odot)>9.9$, of which 452 host an AGN. Redoing our analysis on this sample produced the same results as for the group selection centred on the central galaxy of the corresponding FoF group.

Of final note on the group selection, is the calculation of the halo masses, $M_{200}$. As described in Section \ref{SSG3C}, we calculate our halo masses using the scaling relation of \citet[][see equation \ref{m200eqn}]{Munari2013}, and thus $M_{200}\propto \sigma^{3}$ as would be expected from the virial theorem. Observations using weak gravitational lensing support a scaling relation with a shallower slope, $M_{200} \propto \sigma^{\sim2}$ \citep{Han2015, Viola2015}. This weaker dependency however may not be accurate at the lower halo masses explored in this work. \citet{Viola2015} note that their selection criteria may result in an overestimate of the halo mass for groups with low-velocity, and thus reduce the exponent of the relation. Given this potential bias in the low-mass regime from weak-lensing derived observations, it is prudent for us to use a scaling relation consistent with the virial theorem.

\subsubsection{The infalling population}
Additionally, one must question whether our definition of the infalling and virialised populations may bias our results. We chose to use a physically motivated definition based on the work of \citet{Oman2013}. This was based on $N$-body simulations and the result extrapolated into the projected phase-space plane. An independent test would be to compare our results directly with prior observations, and indeed our radial AGN fractions binned by group halo mass (see Figure \ref{phasespaceR}) support our conclusions in this regard. 

Further support of our results is obtained when we compare directly to \citet{Haines2012} who found no AGN within $0.4R/R_{500}$, and $0.8|\Delta v|/\sigma$ of the cluster centre. Transposing these criteria onto our projected phase-space plane defines a region $R<0.26R_{200}$; $|\Delta v| < 0.8\sigma_\text{group}$. When comparing the fraction of AGN inside this region to that outside we find that at $M_{200}>10^{13.5}\,\text{M}_\odot$ there is a deficit of AGN in this central region at the $2.5\sigma$ level. As with our other results, lower group masses do not show this effect.

\section{Conclusions}
\label{Sconc}
We have exploited the depth and spectroscopic completeness, $98\,$per cent at $r<19.8\,$mag, of the GAMA survey \citep{Driver2011, Liske2015} to probe the halo mass function in order to test the affect of group environment on AGN. We investigated the effect of galaxy position, both radially and within projected phase-space on AGN prevalence. Further to this, by binning our group sample by halo mass, we demonstrate the evolution of a preferential region for AGN fostering within the group projected phase-space. Our main findings are as follows.

\begin{enumerate}
\item The AGN fraction within our group sample is $6.01_{-0.26}^{+0.29}\,$per cent, marginally lower than previous studies have found in groups. We attribute this to the large mass range covered by our sample. Although when split by halo mass the group AGN fraction is approximately flat, for the 25 percent of our galaxies in the lowest mass groups, $M_{200}<10^{13.06}\,\text{M}_\odot$, the AGN fraction is $6.68_{-0.53}^{+0.62}\,$per cent. This is consistent with prior observations of the group AGN fraction \citep{Shen2007, Arnold2009, Oh2014, Tzanavaris2014}.\\
\item We find the AGN fraction as a function of projected separation is flat for low-mass groups. For high mass groups however, the AGN fraction is noticeably reduced at $R<R_{200}$, consistent with the findings of \citet{Ehlert2013, Pimbblet2013} for clusters.\\
\item Using the projected phase-space to split the galaxies into virialised and infalling populations as per \citet[][see equation \ref{OmanEQ}]{Oman2013}, we show that low mass groups do not experience the same deficit of AGN in their cores as do high mass groups and clusters \citep{Gilmour2007, Gavazzi2011, Haines2012}.\\
\item Across the halo mass range of our group sample, the infalling AGN fraction is comparable to the field AGN fraction, indicating that there is no excess of AGN triggering on infall in to groups.\\
\item We demonstrate the evolution of group cores from an environment that supports AGN at low halo masses, to one which inhibits nuclear activity at halo masses greater than $10^{13.5}\,\text{M}_\odot$ (see Figure \ref{MbinnedAGNfrac}).
\end{enumerate}

\section*{Acknowledgements}

The authors take this opportunity to thank the anonymous referee for their constructive comments.
YAG acknowledges the financial support of the University of Hull through an internally funded PhD studentship that has enabled this research to be undertaken.
KAP acknowledge the support of STFC, through the University of Hull's Consolidated Grant ST/R000840/1.
MSO acknowledges the funding support from the Australian Research Council through a Future Fellowship (FT140100255).

GAMA is a joint European-Australasian project based around a spectroscopic campaign using the Anglo-Australian Telescope. The GAMA input catalogue is based on data taken from the Sloan Digital Sky Survey and the UKIRT Infrared Deep Sky Survey. Complementary imaging of the GAMA regions is being obtained by a number of independent survey programmes including GALEX MIS, VST KiDS, VISTA VIKING, WISE, Herschel-ATLAS, GMRT and ASKAP providing UV to radio coverage. GAMA is funded by the STFC (UK), the ARC (Australia), the AAO, and the participating institutions. The GAMA website is \url{http://www.gama-survey.org/}.

Funding for SDSS-III has been provided by the Alfred P. Sloan Foundation, the Participating Institutions, the National Science Foundation, and the U.S. Department of Energy Office of Science. The SDSS-III web site is \url{http://www.sdss3.org/}.

SDSS-III is managed by the Astrophysical Research Consortium for the Participating Institutions of the SDSS-III Collaboration including the University of Arizona, the Brazilian Participation Group, Brookhaven National Laboratory, Carnegie Mellon University, University of Florida, the French Participation Group, the German Participation Group, Harvard University, the Instituto de Astrofisica de Canarias, the Michigan State/Notre Dame/JINA Participation Group, Johns Hopkins University, Lawrence Berkeley National Laboratory, Max Planck Institute for Astrophysics, Max Planck Institute for Extraterrestrial Physics, New Mexico State University, New York University, Ohio State University, Pennsylvania State University, University of Portsmouth, Princeton University, the Spanish Participation Group, University of Tokyo, University of Utah, Vanderbilt University, University of Virginia, University of Washington, and Yale University.

This research made use of Astropy, a community-developed core Python package for Astronomy \citep{AstropyCollaboration2013}.








%
%

\bibliographystyle{mnras}
\bibliography{library} 

\bsp	
\label{lastpage}
\end{document}